\documentclass[11pt]{article}
\def\blue {\color{blue}}
\usepackage{fullpage}
\usepackage{amsfonts}
\usepackage{graphicx}
\usepackage{mathrsfs}
\usepackage{amsmath}
\usepackage{amssymb}
\usepackage{float}
\usepackage{color}
\usepackage[authoryear]{natbib}
\usepackage{subfig}
\usepackage{enumitem}
\usepackage{epstopdf}

\usepackage{geometry}
\newtheorem{thm}{Theorem}
\newtheorem{lemma}{Lemma}
\newtheorem{coro}{Corollary}

\newtheorem{cond}{Condition}

\newtheorem{remark}{Remark}

\def\proof {{\noindent\bf Proof.}\quad}
\def\T{{\top}}
\def\wht{{L}}
\def\bI{\mathbb{I}}
\def\var{\mathrm {Var}}
\def\cov{\mathrm {Cov}}
\def\mB{\mathcal{B}}
\def\L{\mathcal {L}}
\def\mS{\mathcal {S}}
\def\X{{\bf X}}
\def\mC{{\mathcal C}}

\def\A{{\mathcal A}}
\def\mA{\mathcal {A}}

\def\e{{\bf e}}

\def\diag{\mbox{diag}}

\def\D{{\mathcal D}}

\def\In{\tiny\mbox{1}}
\def\out{\tiny\mbox{2}}

\def\bW{{\bf W}}
\def\bA{{\bf A}}

\def\bI{\mathbb{I}}

\def\cp{\mathop{\rightarrow}\limits^{p}}
\newcommand{\bm}{\boldsymbol}
\def\FDR{\mathrm {FDR}}
\def\var{\mathrm {Var}}

\def\cov{\mathrm {Cov}}

\def\FDR{\mathrm {FDR}}
\def\FDP{\mathrm {FDP}}
\def\E{\mathbb {E}}
\def\bH{\mathbb{H}}
\def\hat{\widehat}
\def\mG{\mathcal{G}}

\begin{document}

\title{False Discovery Rate Control Under General Dependence By Symmetrized Data Aggregation}
\date{}

\author{Lilun Du$^1$, Xu Guo$^2$, Wenguang Sun$^3$ and Changliang Zou$^4$}

\maketitle

\begin{abstract}

We develop a new class of distribution--free multiple testing rules for false discovery rate (FDR) control under general dependence. A key element in our proposal is a symmetrized data aggregation (SDA) approach to  incorporating the dependence structure via sample splitting, data screening and information pooling. The proposed SDA filter first constructs a sequence of ranking statistics that fulfill global symmetry properties, and then chooses a data--driven threshold along the ranking to control the FDR. The SDA filter substantially outperforms the knockoff method in power under moderate to strong dependence, and is more robust than existing methods based on asymptotic $p$-values. We first develop finite--sample theories to provide an upper bound for the actual FDR under general dependence, and then establish the asymptotic validity of SDA for both the FDR and false discovery proportion (FDP) control under mild regularity conditions. The procedure is implemented in the R package \texttt{sdafilter}. Numerical results confirm the effectiveness and robustness of SDA in FDR control and show that it achieves substantial power gain over existing methods in many settings.

\medskip
\noindent {\it Keywords}:  Empirical distribution; Integrative multiple testing; Moderate deviation theory; Sample-splitting; Uniform convergence.

\end{abstract}

\footnotetext[1]{Hong Kong University of Science and Technology, Hong Kong}

\footnotetext[2]{Beijing Normal University, Beijing, China}

\footnotetext[3]{University of Southern California. Corresponding Email: wenguans@marshall.usc.edu.}

\footnotetext[4]{Nankai University, Tianjin, China}

\newpage

\section{Introduction}

Multiple testing provides a useful approach to identifying sparse signals from massive data. Recent developments on false discovery rate (FDR; \citealp{benjamini1995controlling}) methodologies have greatly influenced a wide range of scientific disciplines including genomics \citep{Tusetal01, roeder2009genome}, neuroimaging \citep{Pacetal04, Schetal08}, geography \citep{CalSin06, Sunetal15} and finance  \citep{Barras_etal_2010}. Conventional FDR procedures, such as the Benjamini--Hochberg (BH) procedure, adaptive $p$-value  procedure \citep{BenHoc97} and adaptive $z$-value procedure based on local FDR \citep{Efr01, SunCai07}, are developed under the assumption that the test statistics are independent. However, data arising from large--scale testing problems are often dependent. FDR control under dependence is a critical problem that requires much research. Two key issues include (a) how the dependence may affect existing FDR methods, and (b) how to properly incorporate the dependence structure into inference.

\subsection{FDR control under dependence}

The impact of dependence on FDR analysis was first investigated by \cite{benjamini2001control}, who showed that the BH procedure, when adjusted at level $\alpha/(\sum_{j=1}^p 1/j)$ with $p$ being the number of tests, controls the FDR at level $\alpha$ under arbitrary dependence among the $p$-values. However, this adjustment is often too conservative in practice. \cite{benjamini2001control} further proved that applying BH without any adjustment is valid for FDR control for correlated tests satisfying the PRDS property. This result was strengthened by \cite{Sar02}, who showed that the FDR control theory under positive dependence holds for a generalized class of step-wise methods. \cite{storey2004strong}, \cite{Wu08} and \cite{ClaHal09} respectively showed that, in the asymptotic sense,  BH is valid under weak dependence, Markovian dependence and linear process models. Although controlling the FDR does not always require independence, some key quantities in FDR analysis, such as the expectation and variance of the number of false positives, may possess substantially different properties under dependence (\citealp{Owe05, Finetal07}). This implies that conventional FDR methods such as BH can suffer from low power and high variability under strong dependence. \cite{efron2007correlation} and \cite{SchLin11} showed that strong correlations degrade the accuracy in both estimation and testing. In particular, positive/negative correlations can make the empirical null distributions of $z$-values narrower/wider, which has substantial impact on subsequent FDR analyses. These insightful findings suggest that it is crucial to develop new FDR methods tailored to capture the structural information among dependent tests.

Intuitively high correlations can be exploited to aggregate weak signals from individuals to increase the signal to noise ratio (SNR). Hence informative dependence structures can become a bless for FDR analysis. For example, the works of \cite{BenHel07}, \cite{sun2009large} and \cite{SunWei11} showed that incorporating functional, spatial, and temporal correlations into inference can improve the power and interpretability of existing methods. However, these methods are not applicable to general dependence structures.
 \cite{efron2007correlation}, \cite{efron2010correlated} and \cite{fan2012estimating} discussed how to obtain more accurate FDR estimates by taking into account arbitrary dependence. For a general class of dependence models, \cite{leek2008general}, \cite{Frietal09}, \cite{fan2012estimating} and \cite{fan2017estimation} showed that the overall dependence can be much weakened by subtracting the common factors out, and factor--adjusted $p$-values can be employed to construct more powerful FDR procedures. The works by \cite{HalJin10}, \cite{jin2012fdp} and \cite{li2017rate} showed that, under both the global testing and multiple testing contexts, the covariance structures can be utilized, via transformation, to construct test statistics with increased SNR, revealing the beneficial effects of dependence. However, the above methods, for example by \cite{fan2017estimation} and \cite{li2017rate}, rely heavily on the accuracy of estimated models and the asymptotic normality of the test statistics. Under the finite--sample setting, poor estimates of model parameters or violations of normality assumption may lead to less powerful and even invalid FDR procedures. This article aims to develop a robust and assumption--lean method that effectively controls the FDR under general dependence with much improved power.


\subsection{Model and problem formulation}


We consider a setup where $p$-dimensional vectors $\bm\xi_i=(\xi_{i1},\ldots,\xi_{ip})^{\top}$, $i=1, \cdots, n$, follow { a multivariate distribution} with mean $\bm\mu=(\mu_1,\ldots,\mu_p)^{\top}$ and covariance matrix $\bm\Sigma$.
The problem of interest is to test $p$ hypotheses simultaneously:
$$
\mbox{$\bH_j^{0}: \mu_j=0$\; versus\; $\bH_j^{1}: \mu_j \ne 0$,} \quad \mbox{for } j=1, \ldots, p.
$$
The summary statistic $\bar{ \bm\xi}=n^{-1}\sum_{i=1}^n\bm\xi_i$ obeys { a multivariate normal (MVN)} model asymptotically
\begin{equation}\label{mvn}
\bar{ \bm\xi} \overset{d}\approx \mbox{MVN}(\pmb\mu, n^{-1}\bm\Sigma).
\end{equation}
Denote $\bm\Omega=\bm\Sigma^{-1}$ the precision matrix. We first assume that $\bm\Omega$ is known. For the case with unknown precision matrix, a data-driven methodology and its theoretical properties are discussed in Section ~\ref{FDR-asympt2:subsec}.
The problem of multiple testing under dependence can be recast as a variable selection problem in linear regression. Specifically, by taking a ``whitening'' transformation, Model \eqref{mvn} is equivalent to the following model:
\begin{equation}\label{reg-model}
{\bf Y}={\bf X} \pmb\mu+\pmb\epsilon, \quad \pmb\epsilon\overset{d}\approx  \mbox{MVN}(0, n^{-1}{\bf I}_p),
\end{equation}
where ${\bf Y}=\bm\Omega^{1/2}\bar{\bm \xi}\in \mathbb R^p$ is the pseudo response, $\X=\bm\Omega^{1/2}\in \mathbb R^{p\times p}$ is the design matrix, ${\bf I}_p$ is a $p$-dimensional identity matrix and $\pmb\epsilon=(\epsilon_1, \ldots, \epsilon_p)^{\blue \top}$ are noise terms that are approximately independent and normally distributed. The connection between model selection and FDR was discussed in \cite{abramovich2006adapting} and \cite{bogdan2015slope}, respectively under the normal means model and  regression model with orthogonal designs.

Let $\theta_j=\mathbb I\{\mu_j\neq 0\}$, $j=1, \cdots, p$, where $\mathbb I$ is an indicator function, and $\theta_j=0/1$ corresponds to a null/non-null variable. Let $\delta_j\in\{0, 1\}$ be a decision, where $\delta_j=1$ indicates that $\bH_j^0$ is rejected and $\delta_j=0$ otherwise. Let $\mathcal A=\{j: \mu_j\neq 0\}$ denote the non--null set and $\mathcal A^c=\{1, \cdots, p\}\setminus \mathcal A$ the null set. The set of coordinates selected by a multiple testing procedure is denoted $\widehat{\mathcal{A}}=\{j: \delta_j=1\}$. Define the false discovery proportion (FDP) and true discovery proportion (TDP) as:
\begin{equation}\label{FDP}
\mbox{FDP}=\frac{\sum_{j=1}^p (1-\theta_j)\delta_j}{(\sum_{j=1}^p\delta_j)\vee 1}, \quad \mbox{TDP}=\frac{\sum_{j=1}^p \theta_j\delta_j}{(\sum_{j=1}^p\theta_j)\vee 1},
\end{equation}
where $a\vee b=\max(a, b)$. The FDR is the expectation of the FDP: $\mbox{FDR}=\E(\mbox{FDP})$. The average power is defined as $\mbox{AP}=\E(\mbox{TDP})$.

\subsection{FDR control by symmetrized data aggregation}

This article introduces a new information pooling strategy, the symmetrized data aggregation (SDA), for handling the dependence issue in multiple testing. The SDA involves splitting and reassembling data to construct a sequence of statistics fulfilling symmetry properties. Our proposed SDA filter for FDR control consists of three steps:

 \begin{itemize}

  \item The first step splits the sample into two parts, both of which are utilized to construct  statistics to assess the evidence against the null.

  \item The second step aggregates the two statistics to form a new ranking statistic fulfilling symmetry properties.

  \item The third step chooses a threshold along the ranking by exploiting the symmetry property between positive and negative null statistics to control the FDR.

 \end{itemize}

To get intuitions on how the idea works, we start with the independent case [\cite{Zouetal20-t-tests}]. The more interesting but complicated dependent case will be described shortly, with detailed discussions, refinements and justifications deferred to later sections. Suppose the vectors $\bm\xi_i=(\xi_{i1},\ldots,\xi_{ip})^{\top}$ are i.i.d. obeying $\mbox{MVN}(\pmb\mu, {\bf I}_p)$. The proposed SDA method first splits the full sample into two disjoint subsets $\D_1$ and $\D_2$, with sizes $n_1$ and $n_2$ and $n=n_1+n_2$. A pair of statistics, both of which follow $N(0,1)$ under the null, are then calculated to test $\bH_j^0$:
$$
(T_{1j}, T_{2j})=\left\{\frac {\sum_{i\in \D_1} \xi_{ij}}{\sqrt{n_1}}, \frac {\sum_{i\in \D_2} \xi_{ij}}{\sqrt{n_2}}\right\}.
$$
The product $W_j=T_{1j}T_{2j}$ is used to aggregate the evidence across the two groups. If $|\mu_j|$ is large, then both $T_{1j}$ and $T_{2j}$ tend to have large absolute values with the same sign, thereby leading to a positive and large $W_j$. By contrast, $W_j$ fulfills the \emph{symmetry property} under $\bH_j^0$, i.e.
\begin{equation}\label{symmetry:prop}
\mbox{$\Pr(W_j\geq t~|~\bH_j^0)=\Pr(W_j\leq -t~|~\bH_j^0)$, for any $t\in\mathbb{R}.$}
\end{equation}
This motivates one to consider the following selection procedure $\widehat{\mA}=\{j:W_j\geq L\}$, where $L$ is the threshold chosen to control the FDR at level $\alpha$:
\begin{align}\label{th1}
L=\inf\left\{t>0:\frac{\#\{j:W_j\leq -t\}}{\#\{j:W_j\geq t\}\vee 1 }\leq \alpha\right\}.
\end{align}
According to the symmetry property \eqref{symmetry:prop}, the count of negative $W_j$'s below $-t$ strongly resembles the count of false positives in the selected subset (i.e. the null $W_j$'s above $t$). It follows that the fraction in Equation \eqref{th1} provides a good estimate of the FDP.

The dependent case involves a more carefully designed SDA filter. After sample splitting, we apply variable selection techniques such as LASSO to $\D_1$ to construct $T_{1j}$. $T_{1j}$, which is calculated based on linear model \eqref{reg-model}, can effectively capture the dependence structure.
Before using $\D_2$ to construct $T_{2j}$, we carry out a data screening step to narrow down the focus. We show that the screening step can significantly increase the SNR of $T_{2j}$ under strong dependence, hence the correlations are exploited again to increase the power. The ranking statistic $W_j$ is constructed by combining $T_{1j}$ and $T_{2j}$ with proven asymptotic symmetry properties. The theory of the proposed SDA filter is divided into two parts: the finite sample theory provides an upper bound for the FDR  under general dependence, while the asymptotic theory shows that both the FDR and FDP can be controlled at $\alpha+o(1)$  under mild regularity conditions.

\subsection{Connections to existing work and our contributions}

The SDA is closely related to existing ideas of sample--splitting \citep{wasserman2009high, meinshausen2009p} and data carving \citep{Fitetal14, lei2021general}, both of which firstly divide the data into two independent parts,  secondly use one part to narrow down the focus (or rank the hypotheses) and finally use the remainder to perform inference tasks such as variable selection, estimation or multiple testing. These ideas have a common theme with covariate--assisted multiple testing \citep{LeiFit18, Caietal19, LiBar19}, where the primary statistic plays the key role to assess the significance while the side information plays an auxiliary role to assist inference [see also the discussion by \cite{Ram19}]. SDA provides a novel way of data aggregation where both parts of data, which are combined under the symmetry principle, play essential roles in both ranking and selection. This substantially reduces the information loss in conventional sample--splitting methods, while the symmetry principle, which is fulfilled by construction, enables the development of an effective and assumption-lean FDR filter.

The SDA is inspired by the elegant knockoff filter for FDR control \citep{bc2015}, which creates knockoff features that emulate the correlation structure in original features, to form \emph{symmetrized ranking statistics} for selecting important variables via the same mechanism \eqref{th1}. The knockoff method, which is originally developed under regression models, can be applied for FDR control in Model \eqref{mvn} via the equivalent Model \eqref{reg-model}. The knockoff filter employs \emph{local pairwise contrasts}: the ranking variable is constructed to capture the differential evidences against the null exhibited by the pair (i.e. the original feature vs. its knockoff copy). While it is desirable to make the pair as ``independent'' as possible, high correlations will greatly restrict the geometric space in which the knockoff can be constructed; see Appendix~\ref{B.1} for detailed discussions and illustrations. This would significantly increase the difficulty for distinguishing the variable and its knockoff and hence lower the power. By contrast, the SDA filter, which does not rely on pairwise contrasts, will not suffer from high correlations.

\begin{figure}[ht]
\centering
\includegraphics[width=0.8\textwidth]{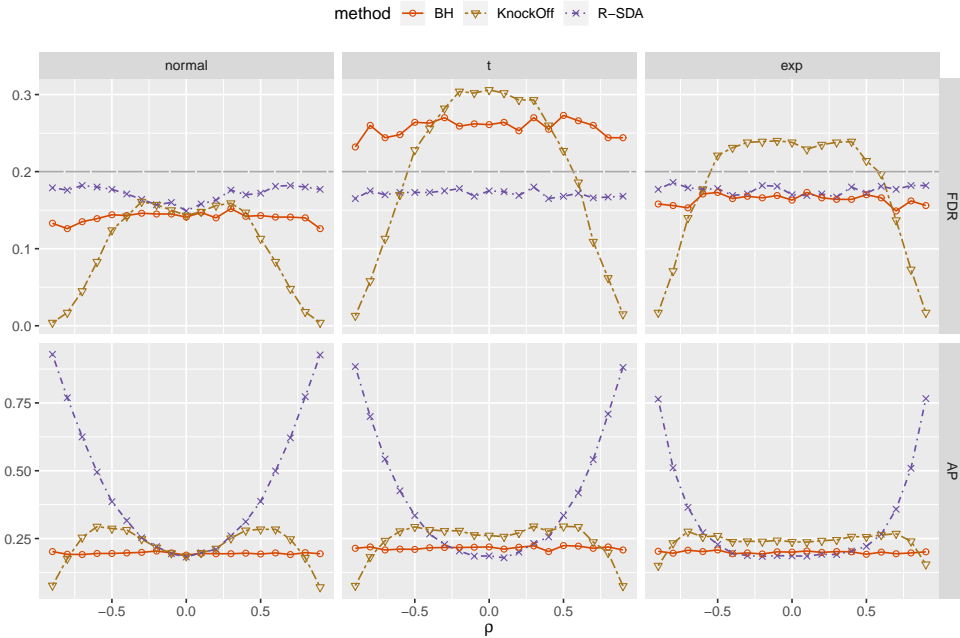}
\vspace{-0.2cm}
\caption {\small \it Impacts of correlation on different FDR procedures: ``$t$'' denotes the $t$ distribution with 3 df and ``exp" denotes
the exponential distribution with scale parameter 2. In both cases the models have been mis-specified as normal when computing the $p$-values.}\label{Fig:intro}
\vspace{-0.5cm}
\end{figure}

To visualize the correlation effects, we consider a setup similar to Figure 5 in \cite{bc2015}, where correlated normal, $t$, and exponential data are generated based on an autoregressive model $\bm\Sigma=(\rho^{|j-i|})$ (see Section \ref{setup:sec} for more details about the setup). We vary $\rho$ from $-0.9$ to $0.9$ and apply BH, knockoff and SDA at FDR level $\alpha=0.2$. The actual FDRs and APs based on 500 replications are summarized in Figure \ref{Fig:intro}. Our first column (normal data) shows that knockoff outperforms BH in some situations, but both the FDR and AP of the knockoff method decrease when correlations grow higher. By contrast, SDA controls the FDR near the nominal level consistently, and the power of SDA increases sharply with growing correlations. This pattern corroborates the insights by \cite{BenHel07}, \cite{sun2009large} and \cite{HalJin10} that high correlations, which can be exploited to increase the SNR, may become a bless in large--scale inference.

The proposed research improves the previous work by \cite{Zouetal20-t-tests} in several ways. First, \cite{Zouetal20-t-tests} has mainly focused on the independent and weak dependent case, with the major goal of deriving convergence rate of false discovery proportions when simultaneously performing thousands of $t$-tests. The methodology in \cite{Zouetal20-t-tests}, which does not utilize LASSO and does not include the data screening step, becomes highly inefficient under strong dependence. See Appendix~\ref{B.2} for an illustration. Second, our new theories for FDR and FDP control \emph{under dependence} and the robustness of the SDA filter under model misspecification substantially depart from the theory in \cite{Zouetal20-t-tests}.

The SDA filter provides a model--free framework that overcomes the limitations of many selective inference procedures, for example, the methods in \cite{locetal14} and \cite{JavJav19}, which require strong assumptions about the conditional distribution to construct asymptotic $p$-values. Our numerical results show that the methods in \cite{fan2017estimation} and \cite{li2017rate}, which require correctly specified models, accurate estimates of parameters and normality assumptions, are in general not robust for FDR control. The SDA filter, which employs empirical distributions instead of asymptotic distributions, only requires the global symmetry of the ranking statistics. It is more robust than its competitors for a wide range of scenarios since the asymptotic symmetry property is much easier to achieve in practice compared to asymptotic normality\footnote{For example, the average of several $t$-variables fulfills the symmetry property perfectly but violates the normality assumption. For asymmetric distributions such as exponential, we usually need a smaller sample size to achieve asymptotic symmetry compared to asymptotic normality -- the latter is stronger than the former since it requires an additional accurate approximation in the tail areas.}. As illustrated by the second column (multivariate $t$ data) of Figure \ref{Fig:intro}, BH fails to control the FDR under heavy--tailed models. The failure in accounting for the deviations from normality may result in misleading empirical null and severe bias in FDR analysis \citep{efron2004large, delaigle2011robustness, liu2014phase}.
{Finally, our Theorem \ref{pro1}, which develops a finite--sample upper bound of FDR under dependence, is closely connected to robust knockoffs
theory and is established utilizing key arguments from \cite{Baretal20}. More specifically, we employ the leave-one-out technique
suggested in \cite{Baretal20} to analyze the effect on the SDA filter of possible
deviations from normality and the sure screening property, similarly to the analysis of the
effect on the Model-X knockoff filter of errors in estimating the true covariance structure.
This important connection sheds lights on how the model uncertainty can affect the actual
FDR level and how the error bound in FDR can be explicitly quantified using appropriate
deviation measures; a detailed discussion is provided in Section~\ref{B.3} of the Supplementary
Material.}

\subsection{Organization}

The remainder of our paper is structured as follows. In Section~\ref{Sec-2}, we introduce the SDA filter for FDR control and discuss the effects of dependence on multiple testing. We develop finite sample and asymptotic theories for FDR control in Section~\ref{Sec-3}. Methodology and theory for the unknown dependence case are discussed in Section~\ref{FDR-asympt2:subsec}.
Simulation and real data analysis are presented in Sections~\ref{Sec-4} and \ref{app:sec}, respectively. The extensions, proofs of theories and additional comparisons are provided in the Supplementary Material.

Notations. For $\mathcal{M}\subset\{1, \cdots, p\}$, let $\bf{X}_{\mathcal{M}}$ be the design matrix
with columns $(\X_{j}: j\in\mathcal{M})$ and {$\X_{j}=(X_{1j},\ldots,X_{pj})^{\top}$} being the $j$th column. For a matrix or a vector ${\bf A}=(a_{ij})$,
$\bA_{\mathcal{M}}$ is similarly defined. Let $\|{\bf A}\|$ be the $L_2$ norm, $\|{\bf A}\|_1=\max_{j}\sum_i|a_{ij}|$, $\|{\bf A}\|_{\max}=\max_{i,j}|a_{ij}|$ and $\|{\bf A}\|_{\infty}=\max_{i}\sum_j|a_{ij}|$. Let $\lambda_{\min}({\bf B})$
and $\lambda_{\max}({\bf B})$ denote the smallest and
largest eigenvalues of a square matrix ${\bf B}$.
The notation $A_n\sim B_n$ means that $A_n/B_n$ and $B_n/A_n$ are both bounded in probability as $n\to\infty$. The ``$\gtrsim$'' and ``$\lesssim$'' are similarly defined.
Let $A_n\approx B_n$ denote the two quantities are asymptotically equivalent, in the sense that $A_n/B_n\cp 1$.

\section{The SDA Filter for FDR Control}\label{Sec-2}

We start with the assumption that the covariance matrix $\bm\Sigma$ is known and then move to the case with unknown $\bm\Sigma$ in Section~\ref{FDR-asympt2:subsec}. Our discussion is mainly based on regression model \eqref{reg-model}; an equivalent description of the methodology via model \eqref{mvn} follows similarly. We first outline in Section \ref{Wj:sec} the steps for constructing the ranking statistics, then provide intuitive explanations on how the SDA filter works in Sections \ref{FDR:sec} and \ref{corr:sec}. The detailed SDA algorithm is provided in Section \ref{sda-algorithm:sec}.

\subsection{Construction of ranking statistics and the symmetry property}\label{Wj:sec}

SDA first splits the data into two independent parts $\D_1$ and $\D_2$, which are respectively used to construct statistics $T_{1j}$ and $T_{2j}$. The information in the two parts is then combined to form the ranking statistic $W_j=T_{1j}T_{2j}$. A wide class of pairs may be constructed from the sample. This section presents a specific pair $(T_{1j}, T_{2j})$, which is used in all numerical studies. Examples of other possible pairs are presented in Section~\ref{other-t1.sec} in the Supplementary Material.

We propose to use LASSO \citep{Tib96} to extract information from $\D_1$ as it simultaneously takes into account the sparsity and dependency structures. Let $\bar{\bm\xi}_1= {n_1}^{-1}\sum_{i\in \mathcal D_1} \bm\xi_i$ and ${\bf y}_1=\X\bar{\bm\xi}_1$. The LASSO estimator is given by $\widehat{\bm\mu}_1=(\hat\mu_{11}, \ldots, \hat\mu_{1p})^\top=\arg\min \L(\bm\mu)$, where
\begin{align}\label{ALPE}
   \L(\bm\mu)=({\bf y}_1-{\bf X}\bm\mu)^{\top}({\bf y}_1-\X\bm\mu)+\lambda\|\bm\mu\|_1.
\end{align}
Let ${\mathcal{S}}=\{j:\widehat{\mu}_{1j}\neq 0\}$ denote the subset of coordinates selected by LASSO and ${\mathcal{S}}^c=\{1, \cdots, p\}\setminus \mathcal S$ its complement.

\begin{remark}\rm{
Following \cite{wasserman2009high}, we suggest using $n_1=\lceil 2n/3\rceil$, which provides stable performance across a wide range of settings. To obtain asymptotically unbiased estimator in the next step, it is required that ${\mathcal{S}}$ contains all the signals with high probability. In practice, this can be achieved by deliberately choosing an overfitted model that includes most true signals and many false positives; see also \cite{BarCan19} and Remark \ref{screening:rem} in Section \ref{FDR-asympt1:subsec}. }
\end{remark}

Next we use $\D_2$ to obtain the least--squares estimates (LSEs). Let $\bar{\bm\xi}_2= {n_2}^{-1}\sum_{i\in \mathcal D_2} \bm\xi_i$, ${\bf y}_2=\X\bar{\bm\xi}_2$, $\X_\mS=(\X_j: j\in \mS)$ and $\e_j=(0,\cdots,0, 1, 0,\cdots, 0)^{\top}$\footnote{Specifically, $\e_j$ is an $|\mS|$-vector with 1 in the $j$th coordinate and 0 elsewhere.}. The LSEs are only calculated for coordinates on the narrowed subset $\mathcal S$. Let $\widehat{\bm\mu}_2=(\hat\mu_{21}, \ldots, \hat\mu_{2p})^\top$, where
\begin{align}\label{lse}
\widehat{\mu}_{2j}=\left\{\begin{array}{cc}\e_j^{\top}({\bf X}_{\mS}^{\top}{\bf X}_{\mS})^{-1}{\bf X}_{\mS}^{\top}{\bf y}_{2}, & \ j\in\mS;\\ 0, & j\in\mS^c. \end{array}	\right.
\end{align}
Section~\ref{corr:sec} provides insights on why this data screening step can lead to increased SNR.

To aggregate information across both $\D_1$ and $\D_2$, let $W_j=T_{1j}T_{2j}$, where
\begin{equation}\label{T1T2}
(T_{1j}, T_{2j})=\left(\frac{ \sqrt{n_1}\hat\mu_{1j}}{\sigma_{\mS,j}}, \frac{\sqrt{n_2}\hat\mu_{2j}}{\sigma_{\mS, j}} \right),
\end{equation}
and ${\sigma}_{\mS, j}^2$'s are the diagonal elements of $(\X_{\mS}^{\T}\X_{\mS})^{-1}$.
A multiple testing procedure consists of two steps: ranking and thresholding. Next we show that $W_j$'s play key roles in both steps.
Intuitively, the positive $W_j$'s can be used for ranking because a large and positive $W_j$ indicates strong evidence against the null. Meanwhile, the negative $W_j$'s, which usually correspond to null cases, can be used for thresholding. The key idea is to exploit the following \emph{asymptotic symmetry property}:
\begin{equation}\label{sym-prop}
\sup_{0 \le t \le c\log p}
\Bigg{|}\frac{\sum_{j\in \mS\cap\mA^c}\bI(W_j \geq t)}{\sum_{j\in \mS\cap\mA^c}\bI(W_j \leq -t)}-1\Bigg{|}=o_p(1) \quad \mbox{for some $c>0$},
\end{equation}
which holds if $P(\mathcal{A}\subseteq\mathcal{S})\rightarrow 1$\footnote{We shall see that ${\mathcal{S}}$ contains all signals, then the LSEs of the null coordinates are symmetrically distributed around 0. Hence $W_j$'s satisfy \eqref{symmetry:prop}. It is easy to see that \eqref{sym-prop} is an asymptotic version of the symmetry property given by \eqref{symmetry:prop}; see Lemmas \ref{klem1}-\ref{klem2} in Section~\ref{Appendix-A} of the Supplementary Material for a rigorous discussion.}. Next we explain how the SDA filter works.

\subsection{FDR thresholding}\label{FDR:sec}

The asymptotic symmetry property \eqref{sym-prop} motivates us to choose the following data--driven threshold to control the FDR at level $\alpha$:
\begin{align}\label{th}
L=\inf\left\{t>0:\frac{\#\{j:W_j\leq -t\}}{\#\{j:W_j\geq t\}\vee 1 }\leq \alpha\right\}.
\end{align}
Our decision rule is given by $\pmb\delta=(\delta_j: 1\leq j\leq p)^\top=\{\mathbb I(W_j\geq L): 1\leq j\leq p\}^\top.$ Denote $\widehat{\mA}=\{j:\delta_j=1\}$ the discovery set. To see why \eqref{th} makes sense, note that $\#\{j: W_{j}\leq -t\}$ is an overestimation of $\#\{j: W_{j}\leq -t, j\in\mA^c\}$, which is asymptotically equal to $\#\{j: W_{j}\geq t, j\in\mA^c\}$, the number of false positives, due to the asymptotic symmetry property \eqref{sym-prop}. It follows that the fraction in \eqref{th}  provides an overestimate of the FDP, which (desirably) leads to a conservative FDR control. Moreover, the empirical FDR level is typically very close to $\alpha$ because the gap between the fraction in \eqref{th} and the actual FDP is usually small in practice, where, for a suitably chosen $L$, most cases in $\{j: W_{j}\leq -L\}$ should come from the null.

\begin{figure}[ht]
\centering
\includegraphics[width=0.8\textwidth]{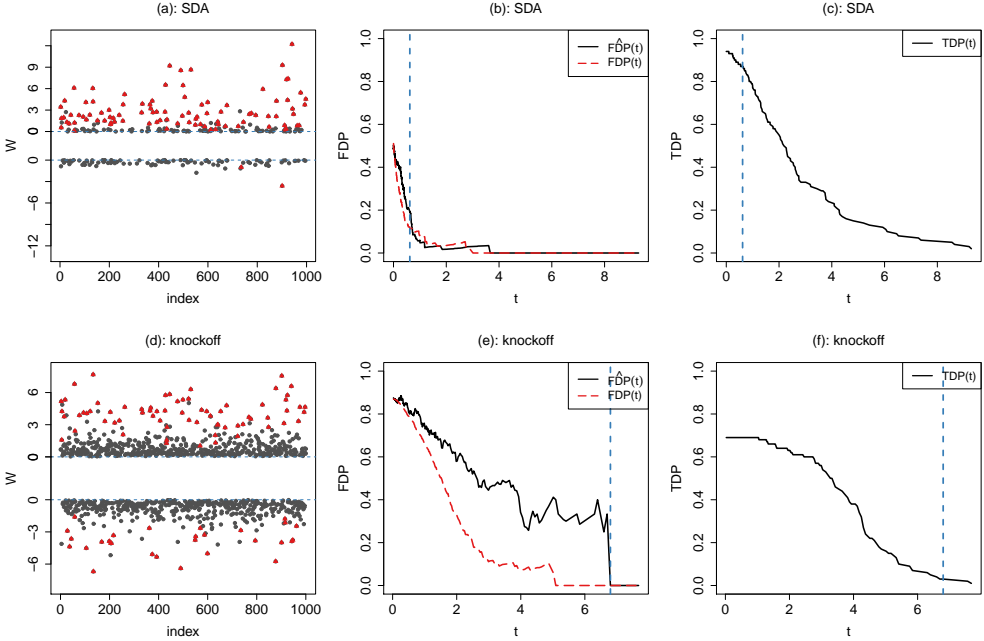}
\vspace{-0.2cm}
\caption {\footnotesize \it (a): Scatter plot of the $288$ nonzero $W_{j}$s from the SDA filter with red triangles and black dots denoting true signals and nulls respectively. A vertical space is added to the middle of the plot to better contrast positive and negative $W_j$'s.
(b): the corresponding estimate of FDP curve (against $t$) along with the true FDP for the SDA filter;
(c): the true power curve (against $t$) for the SDA filter.
(d)-(f): the scatter plot of $p=1000$ $W_j$s, the corresponding FDP estimate, and the true power for the knockoff method.}\label{Fig:tst}
\vspace{-0.2cm}
\end{figure}

The operation of the SDA filter can be visualized in Figure \ref{Fig:tst}. We generate $\{\bm\xi_i: i=1, \ldots, 90\}$ from an MVN distribution with $\bm\mu \in \mathbb{R}^{p=1000}$ and $\bm\Sigma=(0.8^{|i-j|})_{1\leq i, j\leq p}$. We randomly set 10\% of the coordinates in $\pmb\mu$ to be $0.2$ and 0 elsewhere. Panel (a) presents the scatter plot of 288 nonzero $W_{j}$'s with red triangles and black dots respectively denoting true signals and nulls. Panel (d) plots the normalized knockoff statistics that are constructed according to (1.7) in \cite{bc2015}\footnote{The normalization, which makes the plot easier to read, does not affect the results of the knockoff method. This is because only the relative magnitudes of $W_j$ matter in the thresholding step of the knockoff method.}. We can see that both SDA and knockoff fulfill the symmetry property approximately for the null $W_j$'s (black dots). However, SDA achieves a more clearcut separation of signals and noise. As explained in Section \ref{B.1} of the Supplement, the symmetrized knockoff statistics suffers from high correlations.
By contrast, the construction of SDA statistic, which does not depend pairwise contrasts, eliminates the needs for creating fake variables. We can see from Panel (a) that the SDA ranking  places most true signals above 0, and many true signals stay well above the majority of the null cases. However, in Panel (d) that illustrates the knockoff ranking, the true signals are not well separated from the nulls, and many true signals even fall below 0. Since the threshold must be positive, signals with negative $W_j$'s will be missed, which leads to substantial power loss.

The impacts on the FDP processes are shown in the second column in Figure \ref{Fig:tst}. We can see that the estimated FDP process [$\widehat{\mbox{FDP}}(t)$] of SDA approximates the true FDP process [$\mbox{FDP}(t)$] fairly accurately. However, the knockoff method yields overly conservative estimates of the true FDPs, which leads to overly conservative thresholds (marked by blue vertical lines). The last column in Figure \ref{Fig:tst} compares the TDP processes of SDA and knockoff. At the FDR level 0.2, the TDP of SDA is 0.87 ({threshold $L=0.62$}), which is much higher than that of knockoff ({TDP=0.03 with threshold $L=6.80$}). The low TDP of knockoff is due to the decreased power in distinguishing the signal from noise [Panel (d)] and an overly conservative threshold [Panel (e)].

\subsection{Power and effects of dependence}\label{corr:sec}

The impact of dependence on FDR analysis has been extensively studied but most discussions have focused on the validity issue. This section first discusses the impact of dependence on power, and then provides insights on the information loss of conventional data splitting methods.

Under the SDA framework, many possible pairs of $(T_{1j}, T_{2j})$ may be constructed. It is easy to show that $W_j$ constructed via the pairs of sample averages
\begin{equation}\label{t1t2-indep}
(T_{1j}^0, T_{2j}^0)=(\sqrt{n_1}\bar{\bm\xi}_1, \sqrt{n_2}\bar{\bm\xi}_2)
\end{equation}
also fulfill the asymptotic symmetry property. However, the pair in \eqref{t1t2-indep}, which falls into the class of \emph{marginal testing} techniques, can be highly inefficient since it completely ignores the dependence structure. Next we provide intuitions on how the dependence structure is incorporated into the SDA filter to improve the efficiency of existing methods.

First, $T_{1j}$ is superior to $T_{1j}^0$ by leveraging joint modeling techniques. The merit of joint modeling has been carefully illustrated by \cite{bc2015} through extensive simulations. \cite{candes2018panning} further argued that the conditional testing techniques  are in general more powerful in recovering sparse signals than marginal testing methods. $T_{1j}$ is constructed based on LASSO (a conditional inference technique) and serves as a more suitable building block than $T_{1j}^0$ for constructing $W_{1j}$. Second, $T_{2j}$ enjoys a higher SNR than $T_{2j}^0$ by exploiting the dependence between ${{\bm\xi}}_{\mS}$ and ${{\bm\xi}}_{\mS^c}$. Clearly, the expectations of both $\widehat{\bm\mu}_{2\mS}$ and $\bar{\bm\xi}_{2\mS}$ are $\bm\mu_{2\mS}$. The covariance of $\widehat{\bm\mu}_{2\mS}$ is $n_2^{-1}{\bf Q}$, where ${\bf Q}=(\X_{\mathcal{S}}^{\T}\X_{\mathcal{S}})^{-1}$. By the inversion formula of a block matrix, we have
$
\X_{\mathcal{S}}^{\T}\X_{\mathcal{S}}=\bm\Omega_{\mS,\mS}=\left({\bf\Sigma}_{\mS,\mS}-{\bf\Sigma}_{\mS,\mS^c}{\bf\Sigma}_{\mS^c,\mS^c}^{-1}{\bf\Sigma}_{\mS^c,\mS}\right)^{-1}.
$
Hence, ${\bf Q}={\bf\Sigma}_{\mS,\mS}-{\bf\Sigma}_{\mS,\mS^c}{\bf\Sigma}_{\mS^c,\mS^c}^{-1}{\bf\Sigma}_{\mS^c,\mS}$, which is the conditional covariance of ${{\bm\xi}}_{\mS}$ given ${{\bm\xi}}_{\mS^c}$. Let $s_{jl}$ be the $(j,l)$-th element of $\bm\Sigma$. Then $n_2\var(\bar\xi_{2j})=s_{jj}$. However, $n_2\var(\widehat{\mu}_{2j})=s_{jj}-\e_j^\T{\bf\Sigma}_{\mS,\mS^c}{\bf\Sigma}_{\mS^c,\mS^c}^{-1}{\bf\Sigma}_{\mS^c,\mS}\e_j<s_{jj}.$ This provides the key insight on the effect of data screening. In regression terms, strong correlations indicate that a large fraction of variability in the variables in $\mS$ can be explained by the variables in $\mS^c$. The higher the correlations, the more reductions in the uncertainties and hence the higher SNRs. This explains why SDA  becomes more powerful as correlations increase (Figure \ref{Fig:intro}).

Finally, both knockoff and SDA achieve the symmetry property at the expense of possibly reduced SNR: the former increases the dimension of the design matrix by adding noise variables while the latter involves sample splitting. In contrast with the sample splitting method in \cite{wasserman2009high}, where $\mathcal D_1$ is thrown away after model selection, SDA provides a new aggregation strategy: $T_{1j}$ is kept and combined with $T_{2j}$ to form the ranking statistic $W_j$. This substantially reduces the information loss in conventional sample splitting methods.

\subsection{Effects of data screening}

\begin{figure}[ht]
\centering
\includegraphics[width=0.8\textwidth]{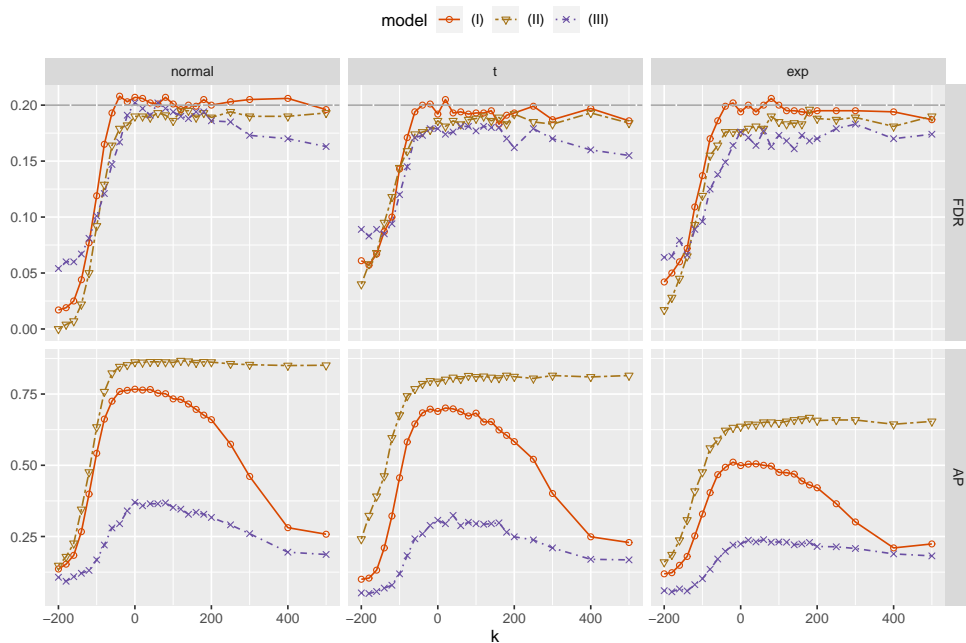}
\vspace{-0.2cm}
\caption {\footnotesize The effects of data screening. We choose $n=90$, $p=500$, and $\mu = \pm 0.2$. The proportion of non-nulls is 10\% and $\alpha =0.2$. We investigate the performance of SDA over 3 distributions and 3 covariance structures described in Section \ref{Sec-4}. Here $k$ denotes the \emph{excess counts} of $|\mathcal S|$ with $\lambda$ selected by the AIC criterion ($k$ can be negative).
}\label{Fig:tst2}
\end{figure}

The data screening step is always beneficial as long as the tests are correlated. Intuitively, the smaller the set $\mS$, the larger amount of uncertainty can be explained by the variables in $\mS^c$. Hence a more effective dimension reduction implies increased SNR and higher power. Meanwhile, our theory on FDR control requires that $P(\mathcal A\subseteq \mathcal S)$ holds with high probability, indicating that an overly aggressive data screening step can hurt the FDR procedure. In practice, we recommend deliberately choosing an overfitted model to ensure the validity in FDR control; this would slightly compromise the power. To illustrate the tradeoff, Figure \ref{Fig:tst2} presents a numerical study to investigate how the size of $\mathcal S$ may affect both the FDR and power. We can see that the actual FDRs of SDA may deviate from the nominal level when $\mathcal S$ is too small. By contrast, a large $\mathcal S$ (overfitted model) has little impact on the FDR levels, but affects the power negatively.

\section{Theoretical Properties of the SDA Filter}\label{Sec-3}

This section first establishes finite sample theory for FDR bounds (Section \ref{FDR-finite:subsec}), and then develops asymptotic theories for FDR and FDP control.

\subsection{Finite--sample theory on FDR control}\label{FDR-finite:subsec}

Our finite--sample theory, which requires no model assumptions, establishes an upper bound for the FDR under general dependence. We emphasize that the upper bound holds for both known and estimated covariance matrices.

Our theory is developed for a modified SDA filter (SDA+) which chooses the threshold
\[
L=\inf\left\{t>0:\frac{1+\#\{j: W_j\leq -t\}}{\#\{j:W_j\geq t\}\vee 1}\leq \alpha\right\}.
\]
SDA+ is slightly more conservative than SDA but their difference is negligible when the number of rejections is large. Recall $\mS=\{j: \hat\mu_{1j}\neq 0\}$. Denote $\bW_\mS=(W_j: j\in\mS)^\T$ and ${\bf W}_{-j}=\bW_{\mS}\setminus W_j$. The key quantity that controls the upper bound is
\begin{equation}
\Delta_j=\left|\Pr(W_{j}>0\mid |W_{j}|,{\bf W}_{-j})-1/2\right|,
\end{equation}
which can be interpreted as a measure of the extent to which the ``flip--sign'' property of $W_j$ is violated\footnote{For a null variable (i.e. $j\in \mA^c$), the flip--sign property means that $W_j$ is equally likely to be positive or negative conditioning on its magnitude and other $W_k$'s in $\mS$.}. Our finite sample theory for FDR control is given by Theorem \ref{pro1}.

\begin{thm}\label{pro1}
For any $\alpha\in(0,1)$, the FDR of the SDA+ method satisfies
\begin{equation}\label{FDR:bound}
\FDR\leq \min_{\epsilon\geq 0}\left\{\alpha(1+5\epsilon)+\Pr\left(\max_{j\in\A^c\cap \mS}\Delta_j>\epsilon\right)\right\}.
\end{equation}
\end{thm}

Our theorem is closely connected to Theorem 1 in \cite{Baretal20}. Both theorems  involve assessing how the deviations from the ``idealized situation'' would affect the  actual FDR level. However, the interpretations are very different. In model-X knockoff the deviation (from the assumption of a known X matrix) comes from the estimation errors of the X matrix whereas in SDA the deviation (from the perfect symmetry property) comes from the possible violations of the normality assumption and sure screening property. Our theorem shows that a tight control of $\Delta_j$'s leads to effective FDR control. Next we carefully interpret the bound and present several important settings in which the upper bound in \eqref{FDR:bound} exactly achieves or is very close to the nominal level $\alpha$.

 Consider the ideal case where (a) the error distribution is symmetric, (b) $\mathcal{S}$ contains all signals and (c) $W_j$'s are independent of each other for $j\in \mathcal{S}$. We can show that $\Delta_j=0$ for all $j\in \mA^c \cap \mS$. The upper bound achieves the nominal level $\alpha$ exactly since $\Pr(W_{j}>0\mid |W_{j}|,{\bf W}_{-j})=\Pr(W_{j}>0\mid |W_{j}|)=1/2$ and hence we can set $\epsilon=0$. Even when the error distribution is asymmetric, we expect that $\Delta_j$'s would become vanishingly small for moderate sample size $n$ due to the convergence of $\widehat{\mu}_{2j}$ to a symmetric distribution (Lemma~\ref{klem1}). Hence the FDR bound would be close to $\alpha$.

Next we turn to the dependent case. For simplicity, assume that $\bm\xi_i$'s come from a multivariate normal distribution. Let ${\bf Q}=(\X_{\mathcal{S}}^{\T}\X_{\mathcal{S}})^{-1}:=(Q_{jk})_{q_n\times q_n}$ with $q_n=|\mathcal{S}|$. The matrix $\bf Q={\bf\Sigma}_{\mS,\mS}-{\bf\Sigma}_{\mS,\mS^c}{\bf\Sigma}_{\mS^c,\mS^c}^{-1}{\bf\Sigma}_{\mS^c,\mS}$ is the conditional covariance matrix of ${{\bm\xi}}_{\mS}$ given ${{\bm\xi}}_{\mS^c}$. The following lemma shows that the magnitude of $\Delta_j$ is controlled by the matrix $\bf Q$.

\begin{lemma}\label{Delta:lem} (Flip--sign property under Gaussian dependence).
Assume that $\bm\xi_i$'s obey a multivariate normal distribution. Denote ${\bf Q}_{-j,j}$ the $j$th column of ${\bf Q}$ excluding $Q_{jj}$. If {${\bf Q}_{-j,j}=\bf 0$}, then $\Delta_j=0$.
\end{lemma}
To provide some intuitions on how close the bound is to $\alpha$ in practice, consider the autoregressive (AR) structure $\bm\Sigma=(\sigma_{j,l})=(\rho^{|j-l|})$. Since the precision matrix of AR structure is tridiagonal, only consecutive coordinates are correlated with each other conditional on remaining variables. Suppose sparse signals are randomly distributed on the $p$ coordinates and the dimension reduction via $\mS$ is performed effectively, e.g. $q_n\ll p$. Let $E$ be an event such that for any null variable $j\in\mS\cap\mA^c$, remaining variables in $\mS$ are conditionally uncorrelated with it. We expect $E$ to occur with high probability since for large tridiagonal precision matrices, there is a small chance that two consecutive coordinates are selected into a small set $\mS$ simultaneously. On event $E$, we have {${\bf Q}_{-j,j}=\bf 0$} and it follows from Lemma \ref{Delta:lem} that $\Delta_j=0$. Consequently the FDR bound would converge to $\alpha$ when $\Pr(E)\rightarrow 1$. In the same vein, we expect that the bound would be close to $\alpha$ for the class of power decay covariance matrices and the class of sparse precision matrices. 


\subsection{Asymptotic theory on FDP control}\label{FDR-asympt1:subsec}

Under the asymptotic paradigm we can prove that the FDR can be controlled at $\alpha+o(1)$ under suitable conditions (\emph{asymptotic validity}).
Denote $\bm\varepsilon_i=\X(\bm\xi_i-\bm\mu)$.
Let $d_n=|\mA|$, $q_n=|\mS|,$ $q_{0n}=|\mS\cap\mA^c|$, and ${\bf A}(\mS):=({\bf X}_{\mS}^\top{\bf X}_{\mS})^{-1}{\bf X}_{\mS}^\top=(a_{jk})_{q_n\times p}$. Assume that $q_{n}$ is uniformly bounded above by some non-random sequence $\bar{q}_n$ that will be specified later. We start with some regularity conditions.

\begin{cond}\label{ssp}
(Sure screening property) As $n\to\infty$, $\Pr(\mathcal{A}\subseteq\mathcal{S})\to 1$.
\end{cond}

\begin{remark}\label{screening:rem}\rm{
Condition \ref{ssp} ensures that $\widehat{\mu}_{2j}$ is unbiased for $j\in \mS$. This pre--selection property, which has been commonly used \citep{wasserman2009high, meinshausen2009p, BarCan19}, can be fulfilled with suitably chosen $\lambda$ under the ``zonal'' assumption \citep{buhlmann2014high}. In practice, we recommend applying AIC to deliberately choose an overfitted model. The sure screening property may not hold exactly but missing small $\mu_j$'s is inconsequential. For example, if we ignore ``unimportant'' signals, then Condition 1 is fulfilled by LASSO for large signals exceeding the rate of $d_n \sqrt{\log p /n}$. Asymptotically unbiased estimators are usually sufficient for effective FDR control. This has been corroborated by our empirical results in Section~\ref{Sec-4}. }
\end{remark}

\begin{cond}\label{lassp}
(Estimation accuracy) The estimator $\widehat{\bm\mu}_1$ fulfills $\|\widehat{\bm\mu}_1-\bm\mu\|_{\infty}=O_p(c_{np})$, where $c_{np}$ is a sequence satisfying $c_{np}\to 0$ and $1/(\sqrt{n}c_{np})=O(1)$.
\end{cond}

\begin{remark}\rm{
Condition \ref{lassp} assumes that $\widehat{\bm\mu}_1$ is a reasonable estimator of $\bm\mu$; this condition typically holds with $c_{np}=d_n\sqrt{\log p/n}$ for the LASSO solution \citep{van2009conditions}. }
\end{remark}

The next two conditions are standard: Condition \ref{moment} imposes constraints on the diverging rates of $\bar{q}_n$ and $p$, both of which depend on the existence of certain moments; Condition \ref{dm} requires that the eigenvalues of the design matrix are doubly bounded by two constants.

\begin{cond}\label{moment}
(Moments) There exist two positive diverging sequences $K_{n1}$ and $K_{n2}$ such that $\mathbb{E}(\|\bm\xi_i-\bm\mu\|_{\infty}^{\theta})\leq K_{n1}^{\theta}$ and
$\mathbb{E}(\|{\bf A}(\mS)\bm\varepsilon_{i}\|_{\infty}^{\theta})\leq K_{n2}^{\theta}$ uniformly in $\mS$ and $i\in\D_2$, where $\theta>2$. Assume that as $n\to\infty$,  $K_{n1}\sqrt{\log p}/{n^{1/2-\gamma-\theta^{-1}}}\to 0$, $\bar{q}_n^{2/\theta}K_{n2}/{n^{1/2-\gamma-\theta^{-1}}}\to 0$ for some small $\gamma>0$.
\end{cond}

\begin{cond}\label{dm}
(Covariance)  There exist positive constants $\bar{\kappa}$ and $\underline{\kappa}$ such that with probability one,
$$\underline{\kappa}\leq \lim\inf_{n\rightarrow\infty}\lambda_{\min}(\X_{\mS}^\T\X_{\mS})<\lim\sup_{n\rightarrow\infty}\lambda_{\max}(\X_{\mS}^\T\X_{\mS})\leq\bar{\kappa}.$$
\end{cond}

\begin{cond}\label{signall}
(Signals)  As $n,p\to\infty$, $\eta_{n} {\color{blue}=} |\mC_{\mu}|\to\infty$, where $$\mathcal{C}_{\mu}=\{j\in\mA: \mu^2_j/\{\max(c^2_{np},\log \bar q_n/n)\}\to\infty\}.$$
\end{cond}

\begin{remark}\rm{
Condition \ref{signall} implies that the number of identifiable effect sizes should not be too small as $p\to\infty$. This seems to be a necessary condition for FDP control. For example, \cite{liu2014phase} showed that if a multiple testing method controls the FDP with high probability, then its number of true alternatives must diverge when the number of tests goes to infinity.}
\end{remark}

\begin{cond}\label{depen}(Dependence) Let $\rho_{jk}=Q_{jk}/\sqrt{Q_{jj} Q_{kk}}$. Assume that for each $j$, $\mbox{Card}\{1\leq k\leq q_n: |\rho_{jk}|\geq C(\log n)^{-2-\nu}\}\leq r_p$, where  $C>0$, $\nu>0$ is any small constant, and $r_p/\eta_n\to 0$ as $n,p\to\infty$.
\end{cond}

\begin{remark}\rm{
Condition  \ref{depen} allows $\xi_j$ to be correlated with all others but requires that the number of large correlations cannot diverge too fast. The condition appears to be similar to the regularity conditions in \cite{fan2012estimating} and \cite{xia2019gap} but in fact our condition is much weaker. For instance, the correlation between $\hat{\mu}_{2 j_1}$ and  $\hat{\mu}_{2 j_2}$ is just the partial correlation of $\xi_{j_1}$ and $\xi_{j_2}$ given the rest variables. In particular, large correlations would be highly unlikely after data screening for a wide range of popular models, such as the class of power decay covariance matrices and the class of moderately sparse precision matrices. This reveals the advantage of SDA, which effectively de--correlates the strong dependence via data screening and conditioning.}
\end{remark}

Our main theoretical result on the asymptotic validity of the SDA method for both FDP and FDR control is given by the next theorem.

\begin{thm}\label{thm1} Suppose Conditions \ref{ssp}-\ref{depen} hold.
For any $\alpha\in(0,1)$, the FDP of the SDA method satisfies
\begin{align}
{\rm FDP}_{W}(L)&:=\frac{\#\{j: W_{j}\geq L,j \in\mA^c\}}{\#\{j: W_{j}\geq L\}\vee 1}\leq\alpha+o_p(1). \label{fa}
\end{align}
It follows that $\mathop{\lim\sup}_{(n,p)\to\infty}{\rm FDR}\leq \alpha$.
\end{thm}


\section{Unknown dependence}\label{FDR-asympt2:subsec}

Now we turn to the case where the covariance structure is unknown. When $\bm\Omega$ is unknown, the SDA filter operates in the same way except that we substitute the estimate $\widehat{\bm \Omega}$ in place of $\bm\Omega$.

We propose to estimate $\bm\Omega$ using \emph{only the first part of the sample} $\D_1$. Denote $\widehat{\bm\Omega}$ the corresponding estimator. Then the SDA filter can be readily constructed via the steps in Sections 2.1-2.2 with ${\bf X}=\widehat{\bm \Omega}^{1/2}$. Various high-dimensional precision matrix estimation methods, such as the graphical LASSO \citep{friedman2008sparse} and CLIME \citep{cai2011constrained}, can be used to obtain $\widehat{\bm\Omega}$. An attractive feature of the SDA filter under unknown dependence is its robustness for FDR control. We next show that the SDA filter is robust for FDR control if $\widehat{\bm\Omega}$ is constructed based only on $\D_1$.
We first state a modified version of Condition 6, which uses ${\bf Q}'$ in place of ${\bf Q}$.

\noindent{\bf Condition 6}' {\it Let ${\bf Q}'=(\X_{\mathcal{S}}^{\T}\X_{\mathcal{S}})^{-1}\X_{\mathcal{S}}^{\T}\X\bm\Omega^{-1}\X^{\T}\X_{\mathcal{S}}(\X_{\mathcal{S}}^{\T}\X_{\mathcal{S}})^{-1}:=(Q_{jk}')_{q_n\times q_n}$ and $\rho_{jk}'=Q_{jk}'/\sqrt{Q_{jj}' Q_{kk}'}$. Assume that for each $j$, $\mbox{Card}\{1\leq k\leq q_n: |\rho_{jk}'|\geq C(\log n)^{-2-\nu}\}\leq r_p$, where $C>0$, $\nu>0$ is any small constant, and $r_p/\eta_n\to 0$ as $n,p\to\infty$.}

The following theorem, which is in parallel with Theorem \ref{thm1}, establishes the asymptotic validity of the SDA filter for estimated covariance.

\begin{thm}\label{corou}
Let $\widehat{\bm\Omega}$ denote an estimator based on $\D_1$. Suppose Conditions \ref{ssp}-\ref{signall} and 6' hold. Then the FDP of the SDA method utilizing $\X=\widehat{\bm\Omega}^{1/2}$ satisfies
 $\mbox{FDP}\leq\alpha+o_p(1)$. It follows that $\mathop{\lim\sup}_{(n,p)\to\infty}{\rm FDR}\leq \alpha$.
\end{thm}

\begin{remark}\rm{
Our FDR theory does not require an accurate estimator for $\bm \Omega$. The accuracy of the estimator only affects the power but not the validity. Consider a \emph{working covariance structure} that ``estimates'' $\bm \Omega$ as the identity matrix. Then it can be shown that the FDP can still be controlled. This is more attractive than the FDR theories in, for example, \cite{fan2017estimation} and \cite{li2017rate} that critically depend on the accuracy of the covariance estimators. }

\end{remark}

The key step in the proof is to verify the validity of (\ref{sym-prop}). This amounts to addressing two major issues: the asymptotic symmetry of $W_j$ under the null and the uniform convergence of $q_{0n}^{-1}\sum_{j\in \mS\cap\mA^c}\bI(W_j \geq t)$. Because $\widehat{\bm\Omega}$ is obtained from $\D_1$, then $\widehat{\mu}_{2j}$ is unbiased conditional on $\D_1$ and thus $\sum_{j \in \mS\cap\mA^c} P(W_j>t)$ is approximately equal to $\sum_{j \in \mS\cap\mA^c} P(W_j<-t)$, establishing the symmetry property. The dependence assumption on ${\bf Q}'$ ensures the convergence of $q_{0n}^{-1}\sum_{j\in \mS\cap\mA^c}\bI(W_j \geq t)$.

While sample--splitting ensures the independence between $\widehat{\bm\mu}_1$ and $\widehat{\bm\mu}_2$ and hence the robustness of the SDA filter, as one would expect, a more accurate estimate of $\bm \Omega$ yields better power. Previously we have proposed to estimate $\bm\Omega$ using $\D_1$ and construct the LSE (\ref{lse}) using $\D_2$. In practice one may consider using $\D_1$ to construct $T_{1j}$, and then obtaining the LSE via the full sample estimator, denoted $\widehat{\bm\Omega}_{F}$, that is estimated using $\{\D_1, \D_2\}$. The caveat is that, although $\X=\widehat{\bm\Omega}_F^{1/2}$ can potentially increase the power, stronger conditions will be needed to guarantee the asymptotic validity of the ``full--sample'' SDA method.
As pointed out by an insightful referee, the asymptotic theory requires that $\widehat{\bm\Omega}_{F}$ must converge to $\bm\Omega$ at a very fast rate, which can be impractical in applications. We recommend the robust SDA filter that estimates $\bm\Omega$ using only $\D_1$. Next we specify the requirements on the estimation accuracy of $\widehat{\bm\Omega}_F$.

\begin{cond}\label{acc} The estimated precision matrix $\widehat{\bm\Omega}_F$ satisfies $\|\widehat{\bm\Omega}_F-{\bm\Omega}\|_{\infty}=O_p(a_{np})$ with $a_{np}\to 0$.
\end{cond}

The following theorem shows that the FDR and FDP can  be controlled asymptotically when $\widehat{\bm\Omega}_{F}$ is sufficiently close to $\bm\Omega$. Let $s_n=\|\bm\Omega\|_{\infty}$.

\begin{thm}\label{coro3}
Consider a modified SDA procedure where we use $\D_1$ to construct $T_{1j}$ and the full sample estimator $\widehat{\bm\Omega}_F$ to construct the LSE (\ref{lse}). Suppose Conditions \ref{ssp}-\ref{depen} hold and $\widehat{\bm\Omega}_{F}$ satisfies Condition \ref{acc}. Then, if
\begin{align}\label{conthm4}
c_{np}a_{np}s_n\bar q_n\sqrt{n\log p}(\log\bar q_n)^{1+\gamma}\to 0
\end{align}
for a small $\gamma>0$, the results in Theorem \ref{thm1} hold for the procedure with $\widehat{\bm\Omega}_{F}$.
\end{thm}

 This theorem, which is a complementary result to Theorem 3, provides conditions that warrant the implementation of a more efficient version of SDA.
It is worth further investigating the condition (\ref{conthm4}), which seems to be unavoidable because $T_{1j}$ and $T_{2j}$ are no longer independent when \emph{the whole sample} is used to estimate $\bm \Omega$. To fix ideas, suppose that ${\bf\Omega}=(\omega_{ij})_{p\times p}$ is $k_n$-sparse, i.e.
$\max_{1\leq i\leq p}\sum_{j\neq i}\mathbb{I}(\omega_{ij}\neq 0)\leq k_n$, and that all its elements $\omega_{ij}$s are bounded. First, standard arguments in, for example, \cite{yuan2010high} and \cite{liu2012high} indicate that $a_{np}=O_p(k_n\sqrt{\log p/n})$. Accordingly, with $c_{np}=d_n\sqrt{\log p/n}$, Equation \eqref{conthm4}
is equivalent to the condition $d_nk_ns_n\bar q_n/n^{1/2}\rightarrow 0$ if $p$ is of a polynomial rate of $n$. The condition above imposes restrictions on the diverging rates of $d_n$, $k_n$, $s_n$ and $\bar q_n$. Assume that $d_n$, $k_n$ and $s_n$ are all bounded. Then we must require that $\bar q_n=o(n^{1/2})$. Alternatively, if we only assume that $k_n$ and $s_n$ are bounded, then a sufficient condition for (\ref{conthm4}) is $\bar q_n=o(n^{1/4})$ (since $d_n\leq \bar q_n$). These rates are consistent with those in the literature; see, for example, \cite{portnoy1984asymptotic} and \cite{fan2004nonconcave}.

\section{Simulation}\label{Sec-4}

This section first introduces the R package \texttt{sdafilter} (Section \ref{pacakge:sec}), followed by simulation designs (Section \ref{setup:sec}) and comparison results (Section \ref{known-cov:sec}). Additional results for comparisons with unknown covariance matrix and other correlation structures are provided in the Supplementary Material.

\subsection{Implementation details} \label{pacakge:sec}

We describe the implementation details of the R package \texttt{sdafilter}. For sample--splitting, we follow the strategy in \cite{wasserman2009high}, which uses $n_1 = [2/3n]$ for selecting variables, and the rest $n_2=n-n_1$ for obtaining the LSEs. The AIC is used to select the tuning parameter in LASSO.  If the number of the variables selected by AIC exceeds $[p/3]$, then only the first $[p/3]$ variables will be retained. For the case with unknown $\bm\Omega$, our default option is to apply the R package \texttt{glasso} to $\D_1$, where the tuning parameter is set by the R package \texttt{huge}. If prior knowledge suggests a nonsparse $\pmb\Omega$, the ``nonsparse'' option in our package can be used. This option first estimates the covariance matrix using the R package \texttt{POET} and then takes its inverse as the input. The \texttt{stable} option implements the R-SDA method described in Section \ref{refine:sec} of the Supplementary Material. {The \texttt{kwd} option enables the usage of different estimators to summarizes the information in the first part of data, including the de-biased LASSO, innovated transformation of the sample means \citep{HalJin10}, and factor-adjusted sample means \citep{fan2017estimation}. }

\subsection{Simulation settings}\label{setup:sec}


We consider three types of covariance structures: (I) Autoregressive (AR) structure: $\bm\Sigma=(\rho^{|j-i|})$. (II) Compound symmetry structure: all off-diagonal elements of the $\bm\Sigma$ are $\rho$, which can be regarded as a factor model with one principal component. (III) Sparse covariance structure: $\bm\Sigma=\bm\Gamma\bm\Gamma^\top+\mathbf{I}_p$, where $\bm\Gamma$ is a $p\times p$ matrix and each row of $\bm\Gamma$ has only one position with nonzero value sampled from uniform distribution $[1, 2]$.

The diagonal elements are normalized as unity for all three settings. To investigate the robustness of different methods, we consider three error distributions: (i) multivariate normal; (ii) $t$-distribution with $\mbox{df}=3$ and (iii) exponential distribution with scale parameter 2. The observations are then standardized to have mean zero and standard deviation one. The correlation structure remains nearly unchanged after transformation. The following six methods will be compared:

\begin{enumerate}
\item [(a)] The Benjamini--Hochberg (BH) procedure with the $p$-values transformed from the $t$ statistics.
\item [(b)] The principal factor approximation (PFA) procedure proposed by \cite{fan2012estimating} for known covariance  and \cite{fan2017estimation} for estimated covariance. Two versions of the PFA procedure using the unadjusted $p$-values and adjusted $p$-values are implemented using the R package \texttt{pfa}, denoted as $\mathrm{PFA}_\mathrm{U}$ and $\mathrm{PFA}_\mathrm{A}$ respectively.
    We only report the results for $\mathrm{PFA}_\mathrm{A}$ as it generally outperforms $\mathrm{PFA}_\mathrm{U}$.
\item [(c)] The sample-splitting method (SS; \citealp{wasserman2009high}), which conducts data screening using LASSO and then applies BH to the $p$-values calculated based on $\widehat{\bm\mu}_2$.
\item [(d)] The knockoff method (Knockoff; \citealp{bc2015}), which is implemented using function ``create.fixed'' in the R package \texttt{knockoff}.
\item [(e)] The DATE method (DATE; \citealp{li2017rate}), which we implemented by ourselves.
\item [(f)] The stability--refined SDA filter (R-SDA) implemented using our package \texttt{sdafilter} with the ``stable'' option. We only presented R-SDA, which we recommend to use in practice, to make the plots easier to read. SDA has similar performance to R-SDA.
\end{enumerate}

Let $n$ be the sample size, $p$ the number tests, and $\pi_1$ the proportion of signals. For each combination $(n, p, \pi_1)$, we generate data and apply the six methods at FDR level $\alpha$. The FDR and AP are calculated by averaging the proportions from 500 replications.


\subsection{Comparison results for known covariance structures}\label{known-cov:sec}

\begin{figure}
  \centering
  \includegraphics[width=0.85\textwidth]{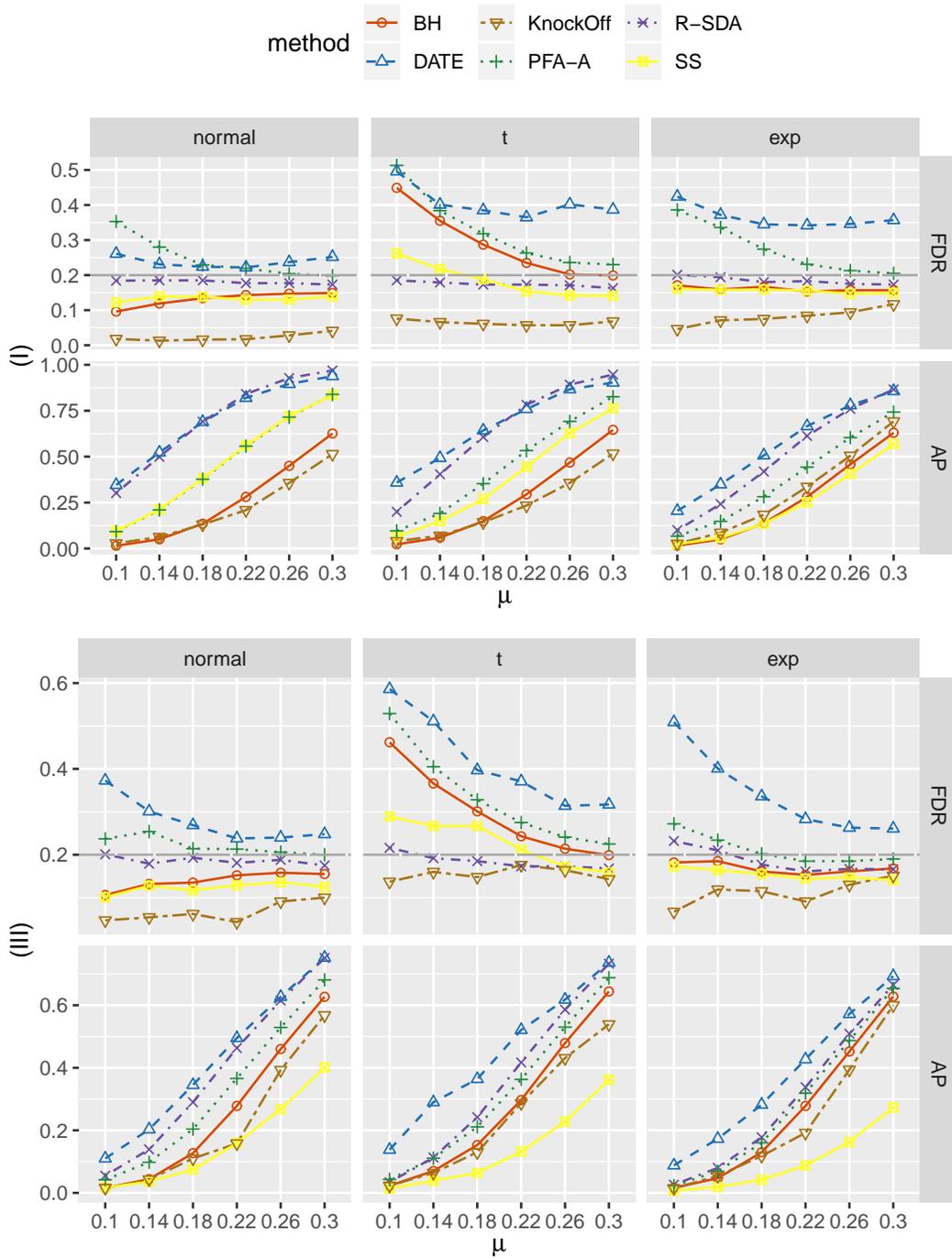}
\vspace{-0.25in}
  \caption{\small\it FDR and AP comparison for varying $\mu$ in Settings (I) and (III) with known variance.}\label{Fig-mu}
\end{figure}

We fix $(n, p, \pi_1, \alpha)=(90, 500, 0.1, 0.2)$ and generate $\mu_j$ from the following random mixture model:
$$
\mu_j \overset{\text{i.i.d.}}{\sim} (1-\pi_1) \delta_0+\pi_1 g(\cdot),  \quad j=1, \cdots, p,
$$
where $\delta_0$ is the dirac delta function (denoting a point mass at 0), and $g(\cdot)$ is the density of the non-null distribution, specified as a uniform distribution $[\mu_0-0.1, \mu_0+0.1]$. { The signals $\mu_j$'s are then randomly multiplied by a flip-sign.} To assess the effect of signal strength, we vary $\mu_0$ from $0.1$ to $0.3$ and apply the six methods to simulated data. The results for Structures (I) and (III) are summarized in Figure~\ref{Fig-mu}, where in the top row we fix $\rho=0.8$. The results for Structure (II) with $\rho=0.8$ are shown in Figure~\ref{Fig-II} of the Supplementary Material. The following observations can be made.
\begin{enumerate}

 \item [(a)] For the Gaussian error case, BH, knockoff, R-SDA and SS control the FDR at the nominal level. The FDR levels of $\mathrm{PFA}_{\mathrm{A}}$ and DATE are inflated when signals are weak.

 \item [(b)] For the non-Gaussian error case, BH, DATE, SS and $\mathrm{PFA}_{\mathrm{A}}$ fail to control the FDR under various settings and the FDR levels can be much higher than the nominal level. Knockoff controls the FDR in all settings but can be very conservative. R-SDA has the most accurate and stable FDR levels among all methods.

  \item [(c)] \emph{R-SDA vs SS and BH.} As expected, SS and BH control the FDR under the Gaussian case but are not robust for non-Gaussian errors. R-SDA has much higher power than both methods (even when the FDR levels of R-SDA are much lower). It is interesting to note that although SS only uses the second part of the data, its power can be much higher than BH when the correlation structure is highly informative [Normal case under Structure (I) on top left]. This is because the data screening step can significantly increase the SNR (Section \ref{corr:sec}).

 \item [(d)] \emph{R-SDA vs Knockoff.} R-SDA and knockoff, both of which are distribution--free, are the only methods that can control the FDR at the nominal level across all scenarios. The knockoff method is overly conservative in Setting (I) due to the high correlation. The conservativeness become less severe under Setting (III). By contrast, R-SDA controls the FDR more accurately near the target level and has significantly higher power than knockoff.

  \item [(e)] \emph{R-SDA vs DATE and $\mathrm{PFA}_\mathrm{A}$.} In some scenarios, DATE and $\mathrm{PFA}_\mathrm{A}$ can outperform SDA in power. However, the higher power may be attributed  to the severely inflated FDRs. The numerical results reveal the promise of extending the SDA framework by employing other methods, such as factor--adjusted $z$-scores or innovated transformations, as alternatives to the LASSO estimates, to construct $T_{1j}$.


\end{enumerate}



\begin{figure}
  \centering
  \includegraphics[width=0.85\textwidth]{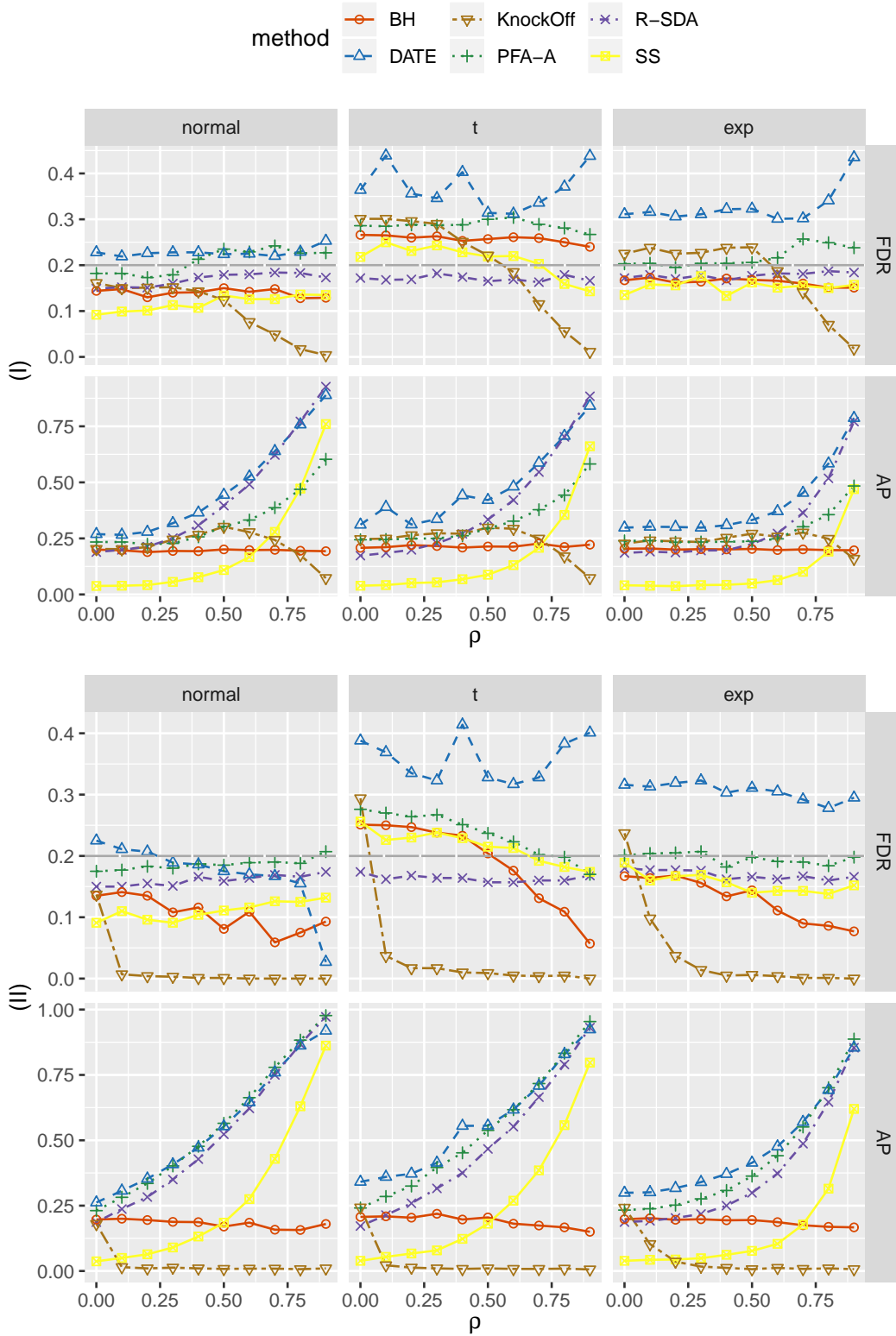}
\vspace{-0.25in}
  \caption{\small\it FDR and AP comparison for varying $\rho$ in Settings (I)--(II) with known covariance matrix.}\label{Fig-rho}
\end{figure}

Next we turn to investigate how the six methods are affected by the strength of correlation. For covariance structures (I) and (II), we fix $\mu=0.2$ under alternative and vary the magnitude of correlation $\rho$ from independence ($\rho=0$) to strong dependence ($\rho=0.9$). The results are summarized in Figure~\ref{Fig-rho}. In addition to the observations that we have made based on the previous graph, the following additional patterns are worthy of mentioning.

\begin{enumerate}

 \item [(a)] The knockoff method becomes more conservative when correlations become higher. Note that the average correlations in Structure (II) is much higher than that in Structure (I), the power of the knockoff method deteriorates faster for Structure (II) as $\rho$ increases. For Structure (II), the FDR of BH also decreases as $\rho$ increases.

 \item [(b)] In contrast with BH and knockoff, both of which suffer from high correlations, the FDR of R-SDA remains at the nominal level consistently, and the power increases with the correlation. The power grows faster for Structure (II).  This corroborates the insights that high correlations can be useful in FDR analysis (\citealp{BenHel07, sun2009large}).

 \item [(c)] In Column 2 of Figure~\ref{Fig-rho}, knockoff fails to control the FDR for heavy tailed distributions when correlation is low. By contrast, SDA controls the FDR accurately under non-Gaussian errors.
\end{enumerate}


\section{A real-data example}\label{app:sec}

This section illustrates the SDA filter for analysis of high-density oligonucleotide microarrays. The data set, which contains $12,625$ probe sets from 128 adult patients enrolled in the Italian GIMEMA multi--center clinical trial, has been used in \cite{chiaretti2005gene} and \cite{Bourgon_etal_2010} for identifying genetic factors that are associated with acute lymphoblastic leukemia (\textsl{ALL}). The \textsl{ALL} dataset is available at {\it http://www.bioconductor.org}.

We focus on a subset of 79 patients with B-cell differentiation because existing research reveals that malignant cells in B-lineage \textsl{ALL} are often associated with genetic abnormalities that have significant impacts on the clinical course of the disease. The patients are divided into two groups based on the molecular heterogeneity of the B-lineage \textsl{ALL}: 37 with the BCR/ABL mutation and 42 with NEG. We further narrow down the focus to 10\% of the genes (i.e., $p=1,263$) before carrying out the FDR analysis. Specifically, the uncorrelated screening method  \citep{Bourgon_etal_2010} has been used to remove probe sets with small overall sample variances since they are unlikely to be differentially expressed.

\begin{figure}
  \centering
  \includegraphics[width=4.5in, height=2.5in]{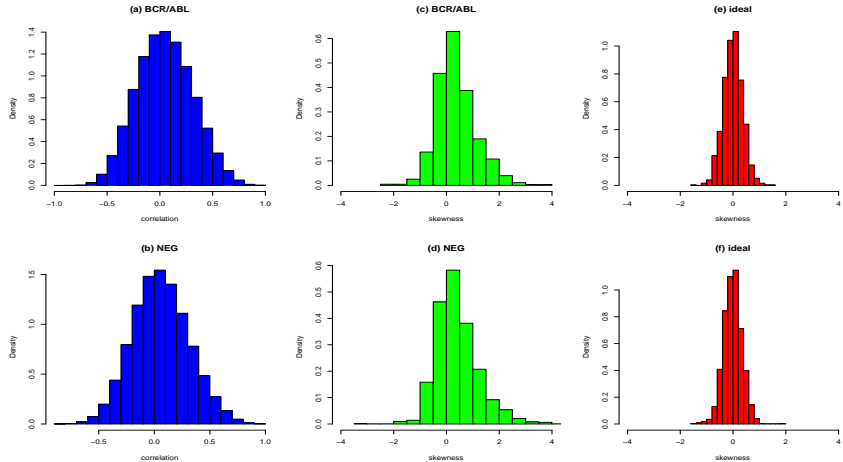}
  \caption{\it
  (a)-(b): Histograms of the off-diagonal elements of the sample correlation matrix for BCR/ABL and NEG;
  (c)-(d): Histogram of the skewness of the $p=1263$ genes for  BCR/ABL and NEG;
  (e)-(f): the ideal patterns of (c)-(d) when the data are normal.
  }\label{Skewness}
\end{figure}

We apply a two--sample version of R-SDA (see Section~\ref{two-sample} for details), BH, SS, $\mathrm{PFA}_{\mathrm{A}}$, Knockoff and DATE at several significance levels for identifying differentially expressed genes across the two groups. Table~\ref{REJ} summarizes the number of significant probe sets for each method. In Figure~\ref{Skewness}(a)-(b), we plot the pairwise correlations of the genes. We can see that a significant proportion of the correlations exceed 0.4. These correlations can jointly exhibit non-negligible dependence effect. This explains why the knockoff method is overly conservative. R-SDA is more powerful than SS by exploiting additional information from the second part of data. BH, $\mathrm{PFA}_{\mathrm{A}}$ and DATE claims more significant genes than R-SDA. However, some caveats need to be given regarding the reliability of BH, $\mathrm{PFA}_{\mathrm{A}}$ and DATE, which all require normality assumptions (and the latter two require accurate estimates of the unknown covariance matrices).

Next we conduct a preliminary analysis to investigate the normality assumption, which seems to have been severely violated in this data set. From Column 2 of Figure~\ref{Skewness} we can see that  the skewness scores of many genes exceed the conventional cutoff $\pm1$. As a comparison, we display in Column 3 of Figure~\ref{Skewness} the ``ideal'' pattern where the normality assumption holds. The histograms in Column 2 are much wider than the histograms in Column 3, indicating a possibly highly skewed error distribution. One possible explanation for the difference in power is that BH, PFA-A and DATE may have inflated FDR levels under violation of normality. This has been observed in our simulation studies (e.g. last column in Figure \ref{Fig-mu-est}). By contrast, SDA and knockoff are distribution--free methods, which tend to produce more reliable and replicable findings. The lists of 19 highest ranked probe sets by the six methods are presented in Table~\ref{Table-genes} of Appendix~\ref{Appendix-C}.

\begin{table}[ht]
\centering \tabcolsep 10.6pt
\caption{\it The number of rejections for six multiple testing procedures and various significance levels.}
\begin{tabular}{c|cccccc}\hline
                         &{\small R-SDA}  & {\small SS}  &{\small BH}   &{\small PFA-A} &{\small Knockoff} &{\small DATE}  \\ \hline
$\alpha=0.01$  &19         &7      &29  &98       &2              &364 \\
$\alpha=0.05$  &33         &15    &146  &182     &2              &452 \\
$\alpha=0.10$  &56         &37    &229  &252     &2              &501 \\
$\alpha=0.20$  &139       &68    &350  &339     &7              &546 \\ \hline
\end{tabular} \label{REJ}
\end{table}

\subsubsection*{Acknowledgments} The authors thank the Editor, Associate Editor and two anonymous referees for their many helpful comments that have resulted in significant
improvements of the article.

{\small \baselineskip 10pt
\bibliographystyle{asa}
\bibliography{references}
}

\newpage

\appendix

\begin{center}\LARGE
Supplementary Material for ``False Discovery Rate Control Under General Dependence By Symmetrized Data Aggregation''
\end{center}

\medskip

\setcounter{equation}{0}
\setcounter{lemma}{0}
\setcounter{figure}{0}
\setcounter{table}{0}
\setcounter{page}{1}

\def\thelemma{S.\arabic{lemma}}
\def\thepro{S.\arabic{pro}}
\def\thelemma{S.\arabic{lemma}}
\def\theequation{S.\arabic{equation}}
\def\thetable{S\arabic{table}}
\def\thefigure{S\arabic{figure}}

This supplement contains some refinements and extensions of the SDA filter (Appendix~\ref{extension:sec}), comparisons of the SDA filter with related ideas in the literature (Appendix~\ref{Comparison}), the proofs of main theorems (Appendix~\ref{Appendix-A}), other theoretical results (Appendix~\ref{Appendix-B}),  and additional numerical results (Appendix~\ref{Appendix-C}).

\section{Refinements and Extensions}\label{extension:sec}

SDA provides a general framework for constructing symmetrized statistics to aggregate structural information from  dependent data. In this section, we discuss some extensions to illustrate how this framework can be implemented in different scenarios.

\subsection{A stability refinement}\label{refine:sec}


To improve the stability in selection and avoid ``$p$-value lottery'' occurred in a single sample splitting \citep{meinshausen2009p}, we propose a modified SDA algorithm that employs the ``bagging'' technique to aggregate results from multiple sample--splitting procedures.

Denote $\widehat{\mA}_k$, $k=1, \ldots, B$, the discovery sets from repeatedly applying $B$ times the SDA filter at level $\alpha$ via random sample splittings. The decisions are aggregated by $\widehat{\mA}_{v}=\#\{j: \sum_{k=1}^B\bI(j\in\widehat{\mA}_k)>\lceil B/2\rceil\}$, the set of variables that are consistently selected in at least 50\% of the replications. The stability refinement picks $\hat\mA_{k^*}$ having the biggest overlap with $\widehat{\mA}_{v}$:
\begin{align}\label{kstar}
k^*=\mathop{\arg\max}_{1\leq k\leq B}\sum_{j=1}^p\left\{\bI(j\in\widehat{\mA}_k\cap\widehat{\mA}_{v})+\bI(j\in\widehat{\mA}_k^c\cap\widehat{\mA}_{v}^c)\right\}.
\end{align}
The new method with stability refinement is denoted R-SDA.
The asymptotic theory for the R-SDA filter is presented and proven in Section~\ref{Appendix-B}.
Our theory implies that the FDPs of $\widehat{\mA}_k$ can be controlled uniformly for all $k$. Hence the discovery set $\widehat{\mA}_{k^*}$ produces more stable results with guaranteed FDR control. Our numerical studies show that compared to SDA,  R-SDA generally yields similar FDR and power but smaller variations in the FDP.

\subsection{Other types of ranking statistics}\label{other-t1.sec}

The SDA filter utilizes $W_j=T_{1j}T_{2j}$ to rank the hypotheses. The asymptotic symmetry property \eqref{sym-prop} is fulfilled as long as $T_{2j}$ are constructed as the LSEs on a subset $\mS$ that includes all signals with high probability. This leaves much flexibility for constructing $T_{1j}$.  We provide a few examples.
\begin{enumerate}[label=\arabic*)]
\item $T_{1j}=\hat\mu_{1j}$, where $\hat\mu_{1j}$ is the LASSO estimate. In contrast with the scaled version  $\hat\mu_{1j}/\sigma_{\mS, j}$, using $\hat\mu_{1j}$ directly reflects the preference of selecting large effect sizes over significant ones. In our numerical studies the two methods seem to perform similarly.

\item If there is prior knowledge that the covariance structure can be well described by a factor model, then we can substitute the factor-adjusted statistics \citep{fan2017estimation} in place of $T_{1j}$.

\item $T_{1j}$ is the de-biased estimate of $\mu_j$ (or its scaled version) based on inverse regression method \citep{xia2019gap}.

\item $T_{1j}$ is the innovated transformation of the sample means \citep{HalJin10, jin2012fdp}.

\end{enumerate}
In our simulation studies, we found LASSO works well and stably in a wide range of settings but can be outperformed by other choices of $T_{1j}$ in special situations. How to develop more powerful ranking statistics is an interesting and challenging problem that requires further research. The main message of this section is that in applications practitioners may develop new types of ranking statistics tailored to problem contexts and prior knowledge about the data structure.

Finally we stress that our theory requires that $T_{2j}$ must be chosen so that the asymptotic symmetry property is fulfilled. For example, it is not allowed to use the LASSO estimate again to construct $T_{2j}$ because this improper choice would lead to a violation of the symmetry property, which no longer guarantees that the FDR can be controlled at the nominal level.

\subsection{Two--sample inference}\label{two-sample}

Suppose we are interested in identifying features that exhibit differential levels across two conditions. Let $\bm\xi^{(k)}=(\xi^{(k)}_{1},\ldots,\xi^{(k)}_{p})^{\top}, k=1, 2,$ be two $p$-dimensional random vectors. The population mean vectors and covariance matrices  are
$\bm\mu^{(k)}$ and $\bm\Sigma^{(k)},k=1,2$, respectively. Consider the following two-sample multiple testing problem:
$$
\mbox{$\bH_j^{0}: \mu^{(1)}_j=\mu^{(2)}_j$ versus $\bH_j^{1}: \mu^{(1)}_j\neq \mu^{(2)}_j$, for $j=1, \ldots, p$.}
$$

The SDA filter can be easily generalized to handle the two-sample situation. Denote $\D^{(k)}=\{\bm\xi^{(k)}_i=(\xi^{(k)}_{i1},\ldots,\xi^{(k)}_{ip})^{\top}, i=1,\cdots, n^{(k)}\}$.  First, we split $\D^{(k)}$ into two disjoint groups $\D^{(k)}_{\In}=(\bm\xi^{(k)}_{\In})$ and $\D^{(k)}_{\out}=(\bm\xi^{(k)}_{\out})$, with
sizes $n^{(k)}_1$ and $n^{(k)}_2$, respectively. Denote $n_l=n_l^{(1)}+n_l^{(2)}, \D_l=\D^{(1)}_{l}\cup \D^{(2)}_{l}, l=1,2$. Based on $\D_{\In}$, the LASSO estimator can be obtained via minimizing
$({\bf y}_1-{\bf X}\bm\omega)^{\top}({\bf y}_1-\X\bm\omega)+\lambda\|\bm\omega\|_1,$ where ${\bf y}_1=\X({\bar{\bm\xi}}^{(1)}_1-{\bar{\bm\xi}}^{(2)}_1)$, $\X=\bm\Omega^{1/2}$, and $\bm\Omega=(n_1/n^{(1)}_1\bm\Sigma^{(1)}+n_1/n^{(2)}_1\bm\Sigma^{(2)})^{-1}$. Denote ${\mathcal{S}}$ the selected subset by  LASSO. Next we calculate the LSEs, using data $\D_2$, for coordinates in $\mS$. The formula is identical to (\ref{lse}) except that now we take ${\bf y}_2=\X({\bar{\bm\xi}}^{(1)}_2-{\bar{\bm\xi}}^{(2)}_2)$ and $\X=\bm\Omega^{1/2}$. Finally, we can calculate $W_j$ and determine the threshold $L$ using (\ref{th}). This procedure is implemented in Section~\ref{app:sec} in the main text to identify differentially expressed genes in microarray studies. Asymptotic theories for the two--sample SDA method, which are presented in Appendix~\ref{Appendix-B}, can be established similarly as done for the standard SDA method.

\subsection{The SDA algorithm: detailed steps}\label{sda-algorithm:sec}

We summarize the operation of the SDA algorithm in this subsection. 

\begin{itemize}

  \item\textit{Step 1:} Split the data set into two parts $\D_1$ and $\D_2$. If the precision matrix $\bm\Omega$ is unknown, use $\D_1$ to obtain its estimate $\widehat{\bm \Omega}$.
  
  \item \textit{Step 2:} Let ${\bf X}=\widehat{\bm \Omega}^{1/2}$. Compute $\widehat{\bm\mu}_1$ by (\ref{ALPE}) and find the narrowed subset $\mathcal S$. Record the estimated coefficients $\hat\mu_{1j}$. 
  
 \item \textit{Step 3:} Compute $\widehat{\bm\mu}_2$ by $(\ref{lse})$ by restricting on the coordinates in the subset $\mathcal S$.
 
 \item \textit{Step 4:}  Compute the ranking statistic $W_j$ by (\ref{T1T2}).
 
 \item \textit{Step 5:} Find the threshold $L$ using (\ref{th}) and output $\widehat{\mathcal{A}}=\{j: W_j\geq L\}$ as the selected features.
 
\end{itemize}

\section{Comparisons with Existing Literature}\label{Comparison}

This section presents comparisons of SDA with existing literature. The goal is to provide insights on the limitations of existing works and highlight some key features of SDA.

\subsection{SDA vs. Knockoff}\label{B.1}

We present some theoretical insights on why the knockoff method suffers from power loss under dependence. The whitening transformation from Model \eqref{mvn} to Model \eqref{reg-model} implies that the fixed-design knockoff filter in \cite{bc2015} is directly applicable to our problem with the Gram matrix ${\bf X}^{\top}{\bf X}=\bm\Omega$, where $\bm\Omega$ is the precision matrix. The augmented design matrix can accordingly be constructed as $(\bm\Omega^{1/2},{\bf 0})^{\top}$ (c.f. Section 2.1.2 of \citealp{bc2015}). The knockoffs $\tilde{\bf X}$ must fulfill $\tilde{\bf X}^{\top}\tilde{\bf X}=\bm\Omega$ and ${\bf X}^{\top}\tilde{\bf X}=\bm\Omega-\mbox{diag}\{ \mathbf{s}\} $, where ${\bf s}=(s_1,\ldots,s_p)^{\top}$ is a $p$-dimensional nonnegative vector. Denote $\mathbf{X}_j$ the $j$th column of the design matrix and $\tilde{\mathbf{X}}_j$ its knockoff copy. In a setting where the features are normalized, i.e. $\Omega_{jj}=1$ for all $j$, the correlation between $\mathbf{X}_j$ and $\tilde{\mathbf{X}}_j$ is $1-s_j$, where $0 \le s_j\le 1$. Intuitively, it is desirable to make the entries of ${\bf s}$ as large as possible; this ensures that $X_j$ would deviate from its knockoff copy as much as possible (hence we will hopefully have sufficient power to distinguish the true signals from faked ones).

\begin{figure}[ht]
\centering
\includegraphics[width=0.9\textwidth]{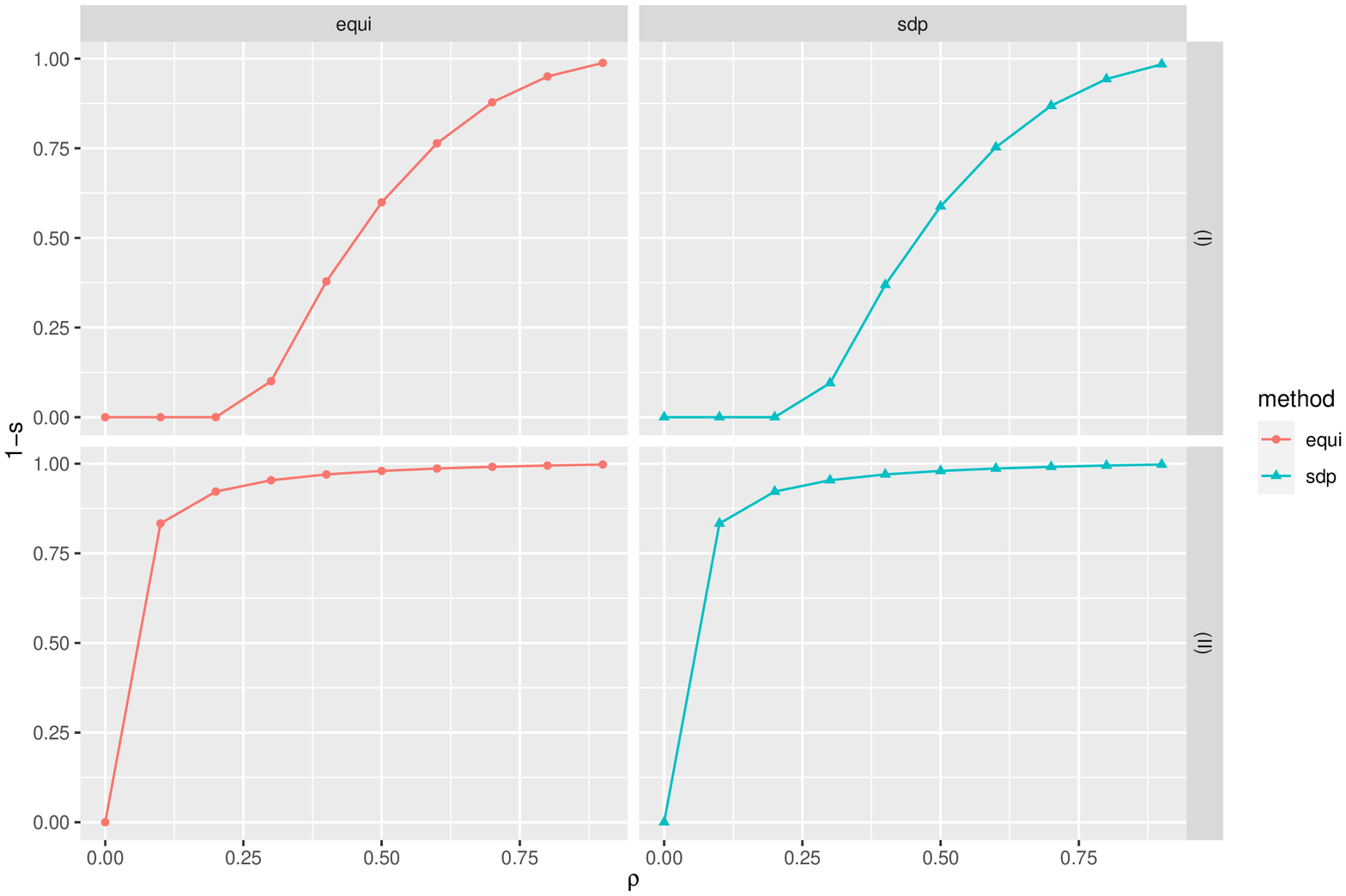}
\vspace{-0.2cm}
\caption {\small \it  The knockoff filter suffers from power loss under moderate to strong dependence. The average similarity score (i.e., $1-s$) between the original variable and its knockoff as a function of $\rho$. Top row: AR(1) structure; bottom row: compound symmetric structure. Both equi-correlated knockoff (left) and SDP knockoff (right) have been  considered. The number of tests is $p=100$.}\label{Fig:knockoff}
\end{figure}

Consider two settings where the correlation structures are respectively AR(1) [$\mbox{Corr}(X_j, X_k)=\rho^{|j-k|}, j\neq k$] and compound symmetric [$\mbox{Corr}(X_j, X_k)=\rho, j\neq k$]. We consider two approaches, namely equi-correlated and SDP knockoffs, both of which were considered in \cite{bc2015} for optimizing $s_j$'s. Figure~\ref{Fig:knockoff} depicts the ``average similarity score'' $1-s$ as a function of different correlation levels $\rho$, where $s=p^{-1}\sum_{j=1}^{p}s_j$ is calculated using both the equi-correlated (left column) and SDP (right column) optimizers. The plots for AR(1) and compound symmetric structures are shown in the top and bottom rows, respectively. We can see that the similarity score $1-s$ increases rapidly in $\rho$. For example,
$1-s$ has already exceeded 95\% when $\rho$ is only $0.25$ under the compound symmetric structure. Consequently, it becomes extremely difficult to distinguish the original variables and their faked copies. This leads to substantial power loss of the knockoff filter. The relationship between the similarity scores and the correlation levels are consistent with the patterns in the power loss of the knockoff method as noted in Fig.5 of \cite{bc2015} and Figure~\ref{Fig:intro} in the main text of this article.

In contrast with the knockoff filter, the operation of SDA does not rely on pairwise contrasts. It only utilizes the \emph{global symmetry property} among all $W_j$'s. The sample-splitting approach eliminates the needs for constructing fake variables under a possibly highly restricted geometric space. This explains why the SDA does not suffer from high correlations.

\subsection{SDA vs. RESS}\label{B.2}

\begin{figure}[ht]
\centering
\includegraphics[width=0.9\textwidth]{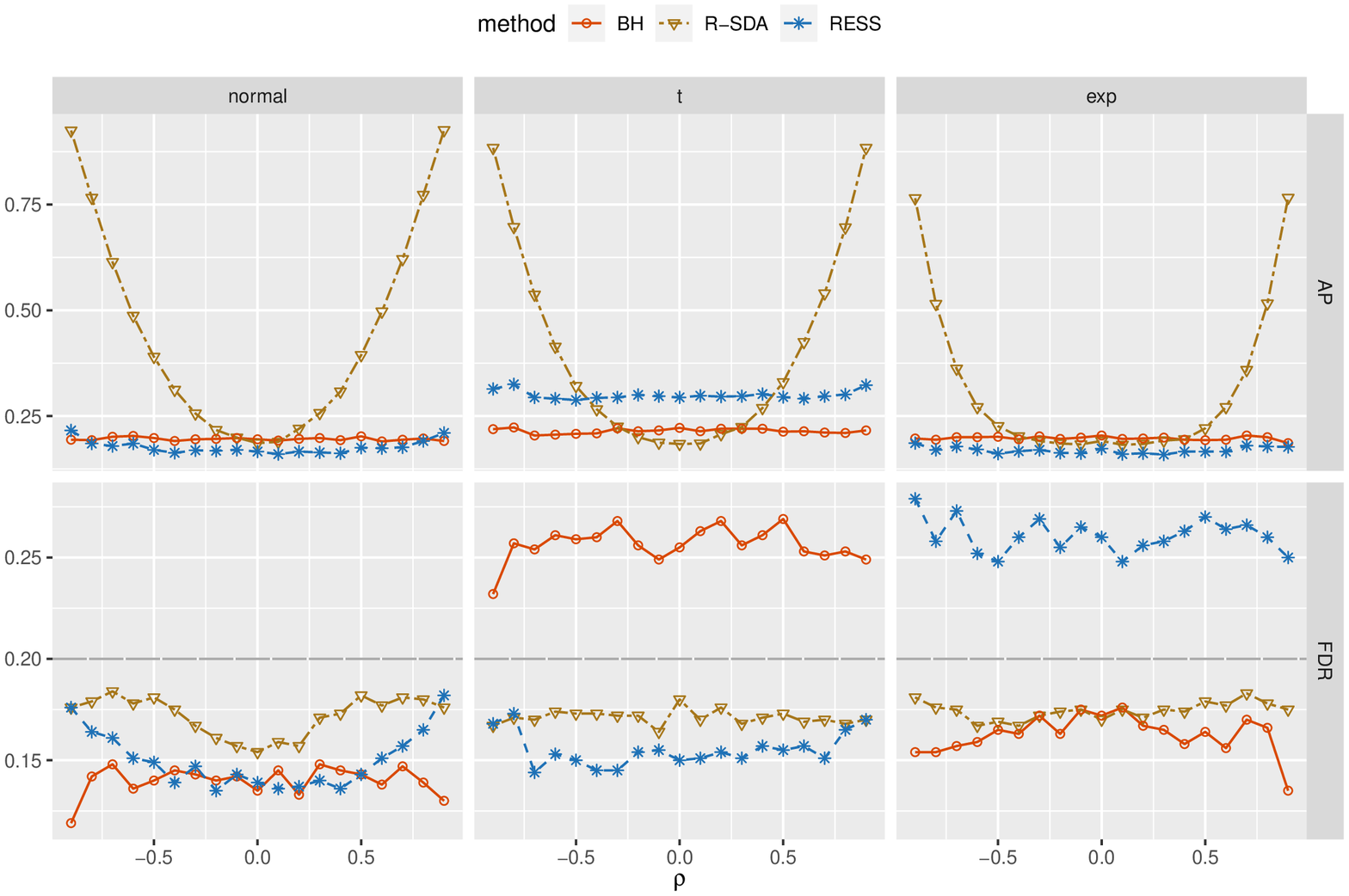}
\vspace{-0.2cm}
\caption {\small \it Impacts of correlation on different FDR procedures. Here $RESS$ refers to the Refection via Sample Splitting procedure in \cite{Zouetal20-t-tests}.}\label{Fig:ttest}
\end{figure}

The reflection via sample-splitting (RESS) method in \cite{Zouetal20-t-tests} was developed for independent two-sample t-tests. It can be substantially improved by SDA that effectively exploits the informative dependence structure. For illustration, Figure~\ref{Fig:ttest} compares the FDR levels and average powers (AP) for SDA vs. BH and RESS in \cite{Zouetal20-t-tests} at different correlation levels. The simulation settings are the same as those in Figure~\ref{Fig:intro} in the main {text}. We can see that the average powers of RESS and BH remain roughly the same across all correlation levels since the dependence structure has been ignored. In contrast, the power of SDA increases sharply with growing correlation levels. Section 2.3 in the main text provides high-level ideas on how the dependence is incorporated into the SDA filter to improve the power.

\medskip
\medskip

\subsection{Model uncertainty and error bound for FDR analysis}\label{B.3}

This section highlights the important connection of our theory to the robust knockoff theory in \cite{Baretal20}, as pointed out by an insightful  referee.

The model-X knockoff assumes that the distribution of the feature vector $X$ is known exactly. However, in practical situations the $X$ distribution must be estimated. In Theorem 1 of \cite{Baretal20}, the KL divergence between the true distribution and its estimate is employed to quantify the effect of estimation errors on FDR control. The KL divergence can be interpreted as a measure of the extent to which the pairwise exchangeability property of the model-X knockoff is violated.

Under the SDA inferential framework, the idealized setting corresponds to the case where the error distribution is perfectly symmetric about 0 and $W_j$'s are independent of each other for $j\in \mathcal S$. This idealized situation implies that $\Pr(W_{j}>0\mid |W_{j}|,{\bf W}_{-j})=1/2$. We call this, borrowing the term from \cite{Baretal20}, the flip-sign property, which indicates that $W_j$ is equally likely to be positive or negative conditional on its magnitude and other $W_k$'s in $\mathcal S$. However, in practical situations the flip-sign property only holds asymptotically. Therefore the actual FDR would unfortunately deviate from the nominal level. The amount of deviation is characterized by
$$
\Delta_j=|\Pr(W_{j}>0\mid |W_{j}|,{\bf W}_{-j})-1/2|,
$$
which can be interpreted as a measure of the extent to which the flip-sign property is violated. We subsequently use $\Delta_j$'s to quantify the effect of asymmetry (i.e. deviation from the perfect symmetry assumption) on FDR control.

\cite{Baretal20} introduced an elegant leave-one-out argument to establish the upper bound for the actual FDR level of the model-X knockoff where the X matrix must be estimated from data. The analysis of SDA in Section \ref{FDR-finite:subsec} reveals that the technique can be readily extended to other important settings where the issue on model uncertainty must be addressed\footnote{In model-X knockoff the model uncertainty comes from the estimation errors whereas in SDA the model uncertainty corresponds to the possible deviation from normality and sure screening property.}. In summary,  the work of \cite{Baretal20} provides a set of useful technical tools for developing finite sample theory on (a) how the FDR control can be affected by the model uncertainty and (b) how the error bound can be explicitly quantified using appropriate deviation measures. The connection of our theory to the robust knockoff theory also provides insights on the impact of deviation from symmetry on the performance of the SDA filter.

\section{Proofs of Main Theorems}\label{Appendix-A}

\subsection{Finite Sample Theory}

This section proves Theorem 1. {The proof of this theorem has extensively used the techniques developed by \cite{Baretal20}, which shows that the Model-X knockoff \citep{candes2018panning} incurs an inflation of the FDR that is proportional to the errors in estimating the distribution of each feature conditional on the remaining features.}

Fix $\epsilon>0$ and for any $t>0$, define
\[
R_{\epsilon}(t)=\frac{\sum_{j\in\mA^c}\bI\left(W_j\geq t,\Delta_j\leq\epsilon\right)}{1+\sum_{j\in\mA^c}\bI\left(W_j\leq -t\right)}.
\]
Consider the event that $\mathcal{A}=\{\Delta:=\max_{j\in\mA^c}\Delta_j\leq\epsilon\}$. Furthermore, consider a thresholding rule $L=T({\bf W})$ that maps statistics ${\bf W}$ to a threshold $L\geq 0$. For each index $j=1,\ldots,p$, {by adopting the leave-one-out argument in \cite{Baretal20},} define
\[
L_j=T\left(W_1,\ldots,W_{j-1},|W_j|,W_{j+1},\ldots,W_{p}\right)\geq 0.
\]

For the SDA filter with threshold $L$, we can write
\begin{align*}
&\frac{\sum_{j\in\mA^c}\bI\left(W_j\geq L,\Delta_j\leq\epsilon\right)}{1\vee\sum_j\bI(W_j\geq L)}=\frac{1+\sum_{j}\bI\left(W_j\leq-L\right)}{1\vee\sum_j\bI(W_j\geq L)} \cdot \frac{\sum_{j\in\mA^c}\bI\left(W_j\geq L,\Delta_j\leq\epsilon\right)}{1+\sum_{j}\bI\left(W_j\leq-L\right)}\\
&\leq \alpha R_{\epsilon}(L).
\end{align*}
Next we derive an upper bound for $\E\{R_{\epsilon}(L)\}$. Note that
\begin{align}
\E\{R_{\epsilon}(L)\}&=\sum_{j\in\mA^c}\E\left\{\frac{\bI\left(W_j\geq L,\Delta_j\leq\epsilon\right)}{1+\sum_{j}\bI\left(W_j\leq -L\right)}\right\} \nonumber \\
&=\sum_{j\in\mA^c}\E\left\{\frac{\bI\left(W_j\geq L_j,\Delta_j\leq\epsilon\right)}{1+\sum_{k\in\mA^c,k\neq j}\bI\left(W_k\leq -L_j\right)}\right\} \nonumber\\
&=\sum_{j\in\mA^c}\E\left[\E\left\{\frac{\bI\left(W_j\geq L_j,\Delta_j\leq\epsilon\right)}{1+\sum_{k\in\mA^c,k\neq j}\bI\left(W_k\leq -L_j\right)}\mid |W_j|,{\bf W}_{-j}\right\}\right]\nonumber \\
&=\sum_{j\in\mA^c}\E\left\{\frac{\Pr\left(W_j>0\mid |W_j|,{\bf W}_{-j}\right)\bI\left(|W_j|\geq L_j,\Delta_j\leq\epsilon\right)}{1+\sum_{k\in\mA^c,k\neq j}\bI\left(W_k\leq -L_j\right)}\right\}. \label{kf}
\end{align}
The last step \eqref{kf} holds since, after conditioning on $(|W_j|,{\bf W}_{-j})$, the only unknown quantity is the sign of $W_j$. By the definition of $\Delta_j$, we have
$\Pr\left(W_j>0\mid |W_j|,{\bf W}_{-j}\right)\leq 1/2+\Delta_j$. Hence,
\begin{align*}
& \E \{R_{\epsilon}(L)\}\\
&\leq \sum_{j\in\mA^c}\E\left\{\frac{(\frac12+\Delta_j)\bI\left(|W_j|\geq L_j,\Delta_j\leq\epsilon\right)}{1+\sum_{k\in\mA^c,k\neq j}\bI\left(W_k\leq -L_j\right)}\right\}\\
&\leq(\frac12+\epsilon)\left[\sum_{j\in\mA^c}\E\left\{\frac{\bI\left(W_j\geq L_j,\Delta_j\leq\epsilon\right)}{1+\sum_{k\in\mA^c,k\neq j}\bI\left(W_k\leq -L_j\right)}\right\}+\sum_{j\in\mA^c}\E\left\{\frac{\bI\left(W_j\leq -L_j\right)}{1+\sum_{k\in\mA^c,k\neq j}\bI\left(W_k\leq -L_j\right)}\right\}\right]\\
&=(\frac12+\epsilon)\left[\E\{R_{\epsilon}(L)\}+\sum_{j\in\mA^c}\E\left\{\frac{\bI\left(W_j\leq -L_j\right)}{1+\sum_{k\in\mA^c,k\neq j}\bI\left(W_k\leq -L_j\right)}\right\}\right].
\end{align*}
The sum in the last expression can be simplified. If for all null $j$, $W_j>-L_j$, then the sum is equal to zero. Otherwise
\begin{align*}
\sum_{j\in\mA^c}\E\left\{\frac{\bI\left(W_j\leq -L_j\right)}{1+\sum_{k\in\mA^c,k\neq j}\bI\left(W_k\leq -L_j\right)}\right\}=\sum_{j\in\mA^c}\E\left\{\frac{\bI\left(W_j\leq -L_j\right)}{1+\sum_{k\in\mA^c,k\neq j}\bI\left(W_k\leq -L_k\right)}\right\}=1,
\end{align*}
where the first equality holds because for any $j,k$, if $W_j\leq-\min(L_j,L_k)$ and $W_k\leq-\min(L_j,L_k)$, then $L_j=L_k$. Accordingly, we have
\[
\E\{R_{\epsilon}(L)\}\leq \frac{1/2+\epsilon}{1/2-\epsilon}\leq 1+5\epsilon,
\]
which proves the theorem. \hfill$\Box$

\subsection{Asymptotic Theory with Known $\bm\Omega$}

We present the proofs of Theorem \ref{thm1} here along with two key lemmas. The lemmas play key roles in our technical arguments and may be of independent interest in their own rights. Other technical lemmas and proofs are provided in Appendix~\ref{Appendix-B}.

For notational convenience, throughout this section, we consider variables that are included in the set $\mathcal{S}$, and suppress ``$j\in\mathcal{S}$'' in all the summations with respect to $j$. Let $\widetilde{\Phi}(x)=1-\Phi(x)$, $G(t)=q_{0n}^{-1}\sum_{j\in \mA^c}\Pr(W_j\geq t\mid \D_{\In})$, $G_{-}(t)=q_{0n}^{-1}\sum_{j\in \mA^c}\Pr(W_j\leq-t\mid \D_{\In})$ and $G^{-1}(y)=\inf\{t\geq 0: G(t)\leq y\}$ for $0\leq y\leq 1$.

The first lemma characterizes the closeness between $G(t)$ and $G_{-}(t)$.
\begin{lemma}\label{klem1} 
Suppose Conditions \ref{ssp}, \ref{moment}, and \ref{dm} hold. We have
\[
\frac{G(t)}{G_{-}(t)}-1\to 0.
\]
uniformly for all $0\leq t\leq G_{-}^{-1}(\alpha\eta_n/q_{0n})$.
\end{lemma}

\proof
Define $b_n=\sigma\sqrt{C\log \bar{q}_n}$ where $C>4$. Denote $T_{kj}=\sqrt{n_k}\widehat{\mu}_{kj}/\sigma_{j}$ for $j=1,\ldots,q_n$ and $\sigma^2=Q_{jj}/\sigma^2_j$.
Observe that
\begin{align*}
&\frac{G(t)}{G_{-}(t)}-1
=\frac{\sum_{j\in\mA^c}\left\{\Pr(T_{1j}T_{2j}\geq t,|T_{2j}|\leq b_n \mid \D_1)-\Pr(T_{1j}T_{2j}\leq -t,|T_{2j}|\leq b_n\mid \D_1)\right\}}{q_{0n}G_{-}({t})}\\
&+\frac{\sum_{j\in\mA^c}\left\{\Pr(T_{1j}T_{2j}\geq t,|T_{2j}|> b_n \mid \D_1)-\Pr(T_{1j}T_{2j}\leq -t,|T_{2j}|> b_n\mid \D_1)\right\}}{q_{0n}G_{-}({t})}\\
:=& \Delta_1+\Delta_2.
\end{align*}

Firstly, for the term $\Delta_2$, by Lemma \ref{lem1} we obtain that
\begin{align*}
\frac{\sum_{j\in\mA^c}\Pr(T_{1j}T_{2j}\geq t,|T_{2j}|> b_n \mid \D_1)}{q_{0n}G_{-}({t})}\leq \frac{\sum_{j\in\mA^c}\Pr(|T_{2j}|>b_n\mid\D_1)}{\alpha\eta_n}
\lesssim \frac{\bar{q}_n\times o(1/\bar{q}_n)}{\eta_n}.
\end{align*}
It follows that $\Delta_2=o(1)$.

By Lemma \ref{mdm}, it can be verified that
\begin{align*}
\frac{\Pr(T_{1j}T_{2j}\geq t, |T_{2j}|\leq b_n\mid \D_1)}{\Pr(T_{1j}Z\geq t, |Z|\leq b_n\mid \D_1)}\to 1,
\end{align*}
where $Z\sim N(0,\sigma^2)$ which is independent of $T_{1j}$. Recall that $$\widehat{\mu}_{2j}/\sigma_j=n_2^{-1}\sum_{i=1}^{n_2}\e^\top_j\left(\X^\top_{\mS}\X_{2\mS}\right)^{-1}\X_{2\mS}^\top \bm\varepsilon_i/\sigma_j:=n^{-1}_2\sum_{i=1}^{n_2}\epsilon_{ij}/\sigma_j.$$
Note that $B_n=n_2\sigma^2$ and
$L_n=B_n^{-\theta/2}\sum_{i=1}^{n_2}\E(|{\epsilon}_{ij}|^{\theta})\leq Cn_2^{1-\theta/2} K_{n2}^{\theta}.
$
We have
$$\{2\log(1/L_n)\}^{1/2}\geq [2\log\{n_2^{\theta/2-1}/(K_{n2}^{\theta})\}]^{1/2}\geq \sqrt{4\log \bar{q}_n},$$
according to Condition \ref{dm}.
The result follows by applying Lemma \ref{mdm}.

Similarly we get
\begin{align*}
\frac{\Pr(T_{1j}T_{2j}\leq -t, |T_{2j}|\leq b_n\mid \D_1)}{\Pr(T_{1j}Z\leq -t, |Z|\leq b_n\mid \D_1)}\to 1.
\end{align*}
Note that
$$
\Pr(T_{1j}Z\leq -t, |Z|\leq b_n\mid \D_1)=\Pr(T_{1j}Z\geq t, |Z|\leq b_n\mid \D_1).
$$
This implies that $\Delta_1=o(1)$, which completes the proof.
\hfill$\Box$

The next lemma establishes the uniform convergence of ${\sum_{j\in \mA^c}\bI(W_j\geq t)}/({q_{0n}G(t)})$.
\begin{lemma}\label{klem2}
Suppose Conditions \ref{moment}, \ref{dm}, and \ref{depen} hold. Then, conditional on $\D_1$, we have
\begin{align}
\sup_{0\leq t\leq G^{-1}(\alpha \eta_n/q_{0n})}\left|\frac{\sum_{j\in \mA^c}\bI(W_j\geq t)}{q_{0n}G(t)}-1\right|=o_p(1),\label{mco1}\\
\sup_{0\leq t\leq G_{-}^{-1}(\alpha \eta_n/q_{0n})}\left|\frac{\sum_{j\in \mA^c}\bI(W_j\leq-t)}{q_{0n}G_{-}(t)}-1\right|=o_p(1).
\label{mco2}
\end{align}
\end{lemma}

\proof We only prove the first formula; the second can be proven similarly. In the proof of Lemma \ref{klem1}, we show that
\[
G(t)=q_{0n}^{-1}\sum_{j\in\mA^c}\Pr(T_{1j}T_{2j}\geq t,|T_{2j}|\leq b_n \mid \D_1)\{1+o(1)\}:=\widetilde{G}(t)\{1+o(1)\}.
\]
Similarly we can show that
\[
q_{0n}^{-1}{\sum_{j\in \mA^c}\bI(W_j\geq t)}=q_{0n}^{-1}{\sum_{j\in \mA^c}\bI(W_j\geq t,|T_{2j}|\leq b_n)}\{1+o_p(1)\}.
\]
Hence, it suffices to show that
\begin{align*}
\sup_{0\leq t\leq G^{-1}(\alpha \eta_n/q_{0n})}\left|\frac{\sum_{j\in \mA^c}\bI(W_j\geq t,|T_{2j}|\leq b_n)}{q_{0n}\widetilde{G}(t)}-1\right|=o_p(1).
\end{align*}
Note that the $\widetilde{G}(t)$ is a decreasing and continuous function. Let $a_p=\alpha\eta_n$, $z_0<z_1<\cdots<z_{h_n}\leq 1$ and $t_i=\widetilde{G}^{-1}(z_i)$, where $z_0=a_p/q_{0n}, z_i=a_p/q_{0n}+b_p\exp(i^{\zeta})/q_{0n}, h_n=\{\log((q_{0n}-a_p)/b_p)\}^{1/\zeta}$ with $b_p/a_p\to 0$ and $0<\zeta<1$. Note that $\widetilde{G}(t_i)/\widetilde{G}(t_{i+1})=1+o(1)$ uniformly in $i$. It is therefore enough to derive the convergence rate of
\[
D_n=\sup_{0\leq i\leq h_n}\left|\frac{\sum_{j\in \mA^c}\left\{\bI(W_j>t_i,|T_{2j}|\leq b_n)-\Pr(W_j>t_i,|T_{2j}|\leq b_n\mid\D_{\In})\right\}}{q_{0n}\widetilde{G}(t_i)}\right|.
\]

Define $\mathcal{M}_j=\{k\in \mA^c:\,\, |\rho_{jk}|\geq C(\log n)^{-2-\nu}\}$, $\mB=\{|T_{2j}|\leq b_n, j\in\mA^c\}$ and
\begin{align*}
D(t)&=\E\left[\left(\sum_{j\in \mA^c}\left\{\bI(W_j>t,|T_{2j}|\leq b_n)-\Pr(W_j>t,|T_{2j}|\leq b_n\mid \D_{\In})\right\}\right)^2\mid \D_{\In}\right]\\
&=\sum_{j\in \mA^c}\sum_{k\in \mA^c}\left\{\Pr(W_j>t, W_k>t\mid\D_{\In},\mB)-\Pr(W_k>t\mid\D_{\In},\mB)\Pr(W_j>t\mid\D_{\In},\mB)\right\}\left\{1+o(1)\right\}.
\end{align*}

Note that
\begin{align*}
D(t)&\leq  r_pq_{0n}G(t)+\sum_{j\in \mA^c}\sum_{k\in \mathcal{M}^c_j}\left\{\Pr(W_k>t, W_j>t\mid\D_{\In},\mB)-\Pr(W_k>t\mid\D_{\In},\mB)\Pr(W_j>t\mid\D_{\In},\mB)\right\}.
\end{align*}
However, for each $j\in \mA^c$ and $k\in\mathcal{M}^c_j$, conditional on $\D_{\In}$,  the Pearson correlation coefficient between $W_j$ and $W_k$ is $\rho_{jk}$. By Lemma 1 in \cite{cai2016large},
\[
\left|\frac{\Pr(W_k>t, W_j>t\mid\D_{\In},\mB)-\Pr(W_k>t\mid\D_{\In},\mB)\Pr(W_j>t\mid\D_{\In},\mB)}{\Pr(W_k>t\mid\D_{\In},\mB)\Pr(W_j>t\mid\D_{\In},\mB)}\right|\leq A_n,
\]
uniformly holds, where $A_n=(\log n)^{-1-\nu_1}$ for $\nu_1=\min(\nu,1/2)$.

From the above results, we can get
\begin{align*}
\Pr(D_n\geq \eta\mid\D_{\In})&\leq \sum_{i=0}^{h_n}
\Pr\left(\left|\frac{\sum_{j\in \mA^c} [\bI(W_j>t_i,|T_{2j}|\leq b_n)-\Pr(W_j>t_i,|T_{2j}|\leq b_n\mid\D_{\In})]}{q_{0n}\widetilde{G}(t_i)}\right|\geq \epsilon\mid\D_{\In}\right)\\
&\leq  \frac{1}{\epsilon^2}\sum_{i=0}^{h_n}\frac{1}{q_{0n}^2\widetilde{G}^2(t_i)}D(t_i)\\
&\leq \frac{1}{\epsilon^2}\left\{r_p\sum_{i=0}^{h_n}\frac{1}{q_{0n}\widetilde{G}(t_i)}+h_nA_n\right\}.
\end{align*}
Moreover, observe that
\begin{eqnarray*}
&&\sum_{i=0}^{h_n}\frac{1}{q_{0n}\widetilde{G}(t_i)}=\frac{1}{a_p}
+\sum_{i=1}^{h_n}\frac{1}{a_p+b_pe^{i^{\zeta}}}\lesssim b_p^{-1}.
\end{eqnarray*}

Finally, note that (a) $\zeta$ can be arbitrarily close to 1 such that $h_n A_n\to 0$, and (b) $b_p$ can be made arbitrarily large as long as $b_p/a_p\to 0$, we conclude that $D_n=o_p(1)$ when $r_p/\eta_n\to 0$. This completes the proof. $\hfill\Box$

In Lemma~\ref{klem1} and Lemma~\ref{klem2}, we have established the symmetry property and uniform consistency for $W_j$'s. Now we are ready to present the proof of Theorem \ref{thm1}.

\paragraph*{Proof of Theorem \ref{thm1}}

By definition, SDA selects the $j$th variable if $W_j\geq \wht$, where
\[
\wht=\inf\left\{t\geq 0: \sum_{j}\bI(W_j\leq-t)\leq \alpha\max\left(\sum_{j}\bI(W_j\geq t),1\right) \right\}.
\]
We need to establish an asymptotic bound for $L$ so that Lemmas \ref{klem1}-\ref{klem2} can be applied.

Let $t^*=G^{-1}_{-}(\alpha\eta_n/q_{0n})$. It follows from Lemma \ref{klem2} that
$$\alpha\eta_n/q_{0n}=G_{-}(t^*)=\frac{1}{q_{0n}}\sum_{j\in\mA^c}\bI(W_j<-t^*)\{1+o(1)\}.$$
On the other hand, for any $j\in\mathcal{C}_{\mu}$, we can show that $\Pr(W_j<t^*, j\in\mathcal{C}_{\mu})\to 0$. In fact, it is straightforward to see that
\begin{align*}
&\Pr\left(W_j<t^*, \ \mbox{for some}\ j\in\mathcal{C}_{\mu}\right)\\
&\leq \eta_n\Pr\left(T_{1j}T_{2j}-\sqrt{n_1n_2}\mu_j^2/\sigma_j^2<t^*-\sqrt{n_1n_2}\mu_j^2/\sigma_j^2\right)\\
&\leq \eta_n\Pr\left(|\mu_j|\left(|\widehat{\mu}_{1j}-\mu_j|+|\widehat{\mu}_{2j}-\mu_j|\right)+|\widehat{\mu}_{1j}-\mu_j||\widehat{\mu}_{2j}-\mu_j|>\mu_j^2-t^*\sigma_j^2/\sqrt{n_1n_2}\right)\to 0.
\end{align*}
To see the last equation, denote $d_j=\mu_j^2-t^*\sigma_j^2/\sqrt{n_1n_2}$. Under Condition \ref{signall}, it follows that $d_j=\mu_j^2\{1+o(1)\}$. We then get
\begin{align*}
&\Pr\left(|\mu_j|\left(|\widehat{\mu}_{1j}-\mu_j|+|\widehat{\mu}_{2j}-\mu_j|\right)+|\widehat{\mu}_{1j}-\mu_j||\widehat{\mu}_{2j}-\mu_j|>d_j\right)\\
&\leq \Pr\left(|\mu_j|\left(|\widehat{\mu}_{1j}-\mu_j|+|\widehat{\mu}_{2j}-\mu_j|\right)>d_j/2\right)+\Pr\left(|\widehat{\mu}_{1j}-\mu_j||\widehat{\mu}_{2j}-\mu_j|>d_j/2\right)=:H_1+H_2.
\end{align*}
Note that $d_j/|\mu_j|=|\mu_j|\{1+o(1)\}$. We observe that
\begin{align*}
H_1&\leq \Pr\left(|\widehat{\mu}_{1j}-\mu_j|>d_j/(4|\mu_j|)\right)
+\Pr\left(|\widehat{\mu}_{2j}-\mu_j|>d_j/(4|\mu_j|)\right),\\
H_2&\leq \Pr\left(|\widehat{\mu}_{1j}-\mu_j|>c_{np}\right)
+\Pr\left(|\widehat{\mu}_{2j}-\mu_j|>C\sqrt{\log \bar q_n/n}\right).
\end{align*}
Then the result follows from Lemmas \ref{lem1} and Condition \ref{lassp}.

Consequently, we have
$\Pr(\sum_{j}\bI(W_j>t^*)\geq \eta_n)\to 1$. We conclude that
$
\sum_{j}\bI(W_j<-t^*)\lesssim \alpha\eta_{n}\leq \alpha\sum_{j}\bI(W_j>t^*)$, and hence $L\lesssim t^*$. By Lemmas \ref{klem1}-\ref{klem2}, we get
\begin{align}\label{fl1}
\frac{\sum_{j\in \mA^c}\bI(W_j\geq \wht)}{\sum_{j\in \mA^c}\bI(W_j\leq-\wht)}-1\to 0.
\end{align}
Next write
\begin{align*}
{\rm FDP}&=\frac{\sum_{j\in\mA^c}\bI\left(W_j\geq \wht\right)}{1\vee\sum_j\bI(W_j\geq \wht)}=\frac{\sum_{j}\bI\left(W_j\leq- \wht \right)}{1\vee\sum_j\bI(W_j\geq \wht)}\times\frac{\sum_{j\in\mA^c}\bI\left(W_j\geq \wht\right)}{\sum_{j}\bI\left(W_j\leq-\wht\right)}\\
&\leq \alpha\times R(\wht).
\end{align*}
Note that $R(\wht)\leq {\sum_{j\in\mA^c}\bI\left(W_j\geq \wht\right)}/{\sum_{j\in\mA^c}\bI\left(W_j\leq -\wht\right)}$, and thus $\mathop{\lim\sup}_{n\to\infty} {\rm FDP}\leq\alpha$  by (\ref{fl1}). Then, for any $\epsilon>0$,
\[
\FDR\leq (1+\epsilon)\alpha R(\wht)+\Pr\left(\FDP\geq (1+\epsilon)\alpha R(\wht)\right),
\]
 which proves the second part of this theorem. \hfill$\Box$



\subsection{Asymptotic Theory with unknown $\bm\Omega$: Proof of Theorems 3 and 4}

\paragraph*{Proof of Theorem \ref{corou}} The proof follows similar lines as those  of  Theorem  \ref{thm1},  except that we now establish {Lemmas \ref{klem1} and \ref{klem2}} under Conditions 1-5 and 6'. Note that Lemma \ref{lem1} still holds under Conditions \ref{ssp}, \ref{moment}, and \ref{dm}. With unknown $\bm\Omega$, conditional on $\D_1$, the Pearson correlation coefficient between
$W_j$ and $W_k$ is changed to $\rho_{jk}'$. The rest of the proof is essentially the same as that of Theorem \ref{thm1} and thus omitted. \hfill$\Box$

\paragraph*{Proof of Theorem 4} To establish this theorem, we consider another SDA procedure with the statistics $\widetilde{W}_j=\sqrt{n_1n_2}\widehat{\mu}_{1j}\widetilde{\mu}_{2j}/\sigma^2_j$, where $\widetilde{\bm\mu}_2$ is the least--squares estimate that uses $\widetilde{\X}={\bm\Omega}^{1/2}$ and $\widetilde{\bf y}_2=\widetilde{\X}\bar{\bm\xi}_2$. We choose a threshold $\widetilde L>0$ by setting
\[
\widetilde L=\inf\left\{t>0:\frac{\#\{j:\widetilde W_j\leq -t\}}{\#\{j:\widetilde W_j\geq t\}\vee 1 }\leq \alpha\right\}.
\]

The proof of this theorem involves a careful investigation of the difference between $W_j$ and $\widetilde W_j$. The main results are summarized by Lemmas \ref{wtw}-\ref{klemsig}.
Define $\mG=\{j: \mu_j=o(c_{np})\}$.

From Lemma {\ref{wtw}}, we have, for any $j$,
\begin{align*}
W_j-\widetilde W_j=\sqrt{n_1n_2}\widehat\mu_{1j}(\widehat\mu_{2j}-\widetilde\mu_{2j})/\sigma^2_j=O_p(n\times s_n\bar{q}_na_{np}\sqrt{\log p/n})\times\{\mu_j+O_p(c_{np})\}.
\end{align*}
Thus for any $j\in\mG$, under condition that $c_{np}a_{np}s_n\bar{q}_n\sqrt{n\log p}(\log\bar q_n)^{1+\gamma}\to 0$ for a small $\gamma>0$, the absolute difference between $W_j$ and $\widetilde W_j$ is negligible.
While for $j\in\mG^{c}$, we need to consider the relative difference. That is,
\begin{align*}
\widetilde{W}_j=W_j\left\{1+\frac{\widetilde\mu_{2j}-\widehat\mu_{2j}}{\widehat\mu_{2j}}\right\}
=W_j\left\{1+\frac{O_p(s_n\bar{q}_na_{np}\sqrt{\log p/n})}{\mu_j+O_p(\sqrt{\log\bar q_n/n})}\right\}=W_j\{1+o_p(1)\}.
\end{align*}
In fact, under conditions $c_{np}a_{np}s_n\bar{q}_n\sqrt{n\log p}(\log\bar q_n)^{1+\gamma}\to 0$ and $1/(\sqrt n c_{np})=O(1)$, we have:
$$\frac{s_n\bar{q}_na_{np}\sqrt{\log p/n}}{c_{np}}=o(1),\,\,\,\frac{s_n\bar{q}_na_{np}\sqrt{\log p/n}}{\sqrt{\log\bar q_n/n}}=o(1).$$

From Lemma \ref{klem0} and Lemma \ref{klemsig} given below, we conclude that
$${\rm FDP}_{\widetilde{W}}(\widetilde L):=\frac{\#\{j: \widetilde W_{j}\geq \widetilde L,j \in\A^c\}}{\#\{j: \widetilde W_{j}\geq \widetilde L\}\vee 1}={\rm FDP}_W(L)\left\{1+o_p(1)\right\}.$$
Under Conditions 1-6, similar to the proof of Theorem 2, we can show that ${\rm FDP}_{\widetilde{W}}(\widetilde L)$ is controlled at the nominal level asymptotically. Thus the claimed result follows.
\hfill$\Box$

\begin{lemma}\label{wtw}
If  Conditions \ref{ssp}, \ref{moment}, \ref{dm} and \ref{acc} hold, then we have  $\widehat\mu_j=\widetilde{\mu}_{j}+O_p(a_{np}s_n\bar{q}_n\sqrt{\log p/n})$ uniformly in $j\in\mS$.
\end{lemma}

\proof Note that
\begin{align*}
|\widehat\mu_j-\widetilde\mu_j|&=\left|{\bf e}_j^{\T}\left\{(\widetilde{\X}^\top_{\mS}\widetilde{\X}_{\mS})^{-1}\widetilde{\X}^\top_{\mS}\widetilde{\X}-({\X}^\top_{\mS}{\X}_{\mS})^{-1}{\X}^\top_{\mS}{\X}\right\}(\bar{\bm\xi}-\bm\mu)\right|\\
&\leq \left\|(\widetilde{\X}^\top_{\mS}\widetilde{\X}_{\mS})^{-1}\widetilde{\X}^\top_{\mS}\widetilde{\X}-({\X}^\top_{\mS}{\X}_{\mS})^{-1}{\X}^\top_{\mS}{\X}\right\|_{\infty}\|\bar{\bm\xi}-\bm\mu\|_{\infty}\\
&=: \|{\bm\Delta}\|_{\infty}\|\bar{\bm\xi}-\bm\mu\|_{\infty}.
\end{align*}
Similar to Lemma \ref{lem1}, we get $\|\bar{\bm\xi}-\bm\mu\|_{\infty}=O_p(\sqrt{\log p/n})$. For the analysis of $\bm\Delta$, we note the following fact
\begin{align*}
 &\|\widetilde{\X}^\top_{\mS}\widetilde{\X}_{\mS}-{\X}^\top_{\mS}{\X}_{\mS}\|_{\infty}\leq\|\widehat{\bm\Omega}_F-{\bm\Omega}\|_{\infty}=O_p(a_{np}),\\
 &\|\widetilde{\X}^\top_{\mS}\widetilde{\X}_{\mS^c}-{\X}^\top_{\mS}{\X}_{\mS^c}\|_{\infty}\leq\|\widehat{\bm\Omega}_F-{\bm\Omega}\|_{\infty}=O_p(a_{np}),\\
 &\|(\widetilde{\X}^\top_{\mS}\widetilde{\X}_{\mS})^{-1}-({\X}^\top_{\mS}{\X}_{\mS})^{-1}\|_{\infty}\leq\|(\widetilde{\X}^\top_{\mS}\widetilde{\X}_{\mS})^{-1}\|_{\infty}\|({\X}^\top_{\mS}{\X}_{\mS})^{-1}\|_{\infty}\|\widetilde{\X}^\top_{\mS}\widetilde{\X}_{\mS}-{\X}^\top_{\mS}{\X}_{\mS}\|_{\infty}=O_p(\bar{q}_na_{np}).
\end{align*}
Thus, by triangle inequality, we can conclude that
\begin{align*}
\|\bm\Delta\|_{\infty}&=\|(\widetilde{\X}^\top_{\mS}\widetilde{\X}_{\mS})^{-1}\widetilde{\X}^\top_{\mS}\widetilde{\X}_{\mS^c}-({\X}^\top_{\mS}{\X}_{\mS})^{-1}\X_{\mS}^{\top}\X_{\mS^c}\|_{\infty}\\
&\leq \|(\widetilde{\X}^\top_{\mS}\widetilde{\X}_{\mS})^{-1}-({\X}^\top_{\mS}{\X}_{\mS})^{-1}\|_{\infty}\|{\X}^\top_{\mS}{\X}_{\mS^c}\|_{\infty}+\|\widetilde{\X}^\top_{\mS}\widetilde{\X}_{\mS^c}-{\X}^\top_{\mS}{\X}_{\mS^c}\|_{\infty}\|(\widetilde{\X}^\top_{\mS}\widetilde{\X}_{\mS})^{-1}\|_{\infty}\\
&=O_p(\bar{q}_ns_na_{np})
\end{align*}
and accordingly
$\max_j|\widehat\mu_j-\widetilde\mu_j|=O_p(\bar{q}_ns_na_{np}\sqrt{\log p/n})$.
\hfill$\Box$

The next lemma establishes the approximation result of $W_j$ to $\widetilde W_j$ for those $j\in\mG$.
\begin{lemma}\label{klem0}
Suppose Conditions \ref{ssp}, \ref{lassp}, \ref{moment}, \ref{dm} and \ref{acc} hold and \\$c_{np}a_{np}s_n\bar{q}_n\sqrt{n\log p}(\log\bar q_n)^{1+\gamma}\to 0$ for a small $\gamma>0$. Then, for any $M>0$,
\begin{align*}
\sup_{M\leq t\leq G^{-1}(\alpha \eta_n/q_{0n})}\left|\frac{\sum_{j\in \mG}\bI(\widetilde W_{j}\geq t)}{\sum_{j\in \mG}\bI({W}_{j}\geq t)}-1\right|&=o_p(1),\\
\sup_{M\leq t\leq G_{-}^{-1}(\alpha \eta_n/q_{0n})}\left|\frac{\sum_{j\in \mG}\bI(\widetilde W_{j}\leq -t)}{\sum_{j\in \mG}\bI({W}_{j}\leq -t)}-1\right|&=o_p(1).
\end{align*}
\end{lemma}

\proof  By Lemma \ref{wtw}, with probability tending to one,
\begin{align*}
&\left|{\sum_{j\in \mG}\bI(W_{j}\geq t)}-{\sum_{j\in \mG}\bI(\widetilde{W}_{j}\geq t)}\right|\\
&\leq \left|\sum_{j\in \mG}\left\{\bI({W}_{j}\geq t+l_n)-\bI({W}_{j}\geq t)\right\}\right|+\left|\sum_{j\in \mG}\left\{\bI({W}_{j}\geq t-l_n)-\bI({W}_{j}\geq t)\right\}\right|\\
&:=\Delta_1+\Delta_2,
\end{align*}
where $l_n/(c_{np}a_{np}s_n\bar{q}_n\sqrt{n\log p})\to\infty$ as $n,p\to\infty$.
We will deal with $\Delta_1$ only and the part of $\Delta_2$ is similar.  Define the events $\mC_t=\{|T_{1j}|>t/(C\sqrt{\log \bar{q}_n}), |T_{2j}|>t/(\sqrt{n}c_{np}),j\in\mG\}$.
\begin{align*}
\E(\Delta_1)&=\E\left\{\sum_{j\in \mG}\bI(t\leq{W}_{j}\leq t+l_n)\right\}\\
&\leq\sum_{j\in \mG}\Pr(t\leq{W}_{j}\leq t+l_n\mid \mC_t)+\sum_{j\in \mG}\Pr(t\leq{W}_{j}\leq t+l_n,\mC^c_t)\\
&\leq\sum_{j\in \mG}\Pr(t\leq{W}_{j}\leq t+l_n\mid \mC_t)+o(1),
\end{align*}
where we use Lemmas \ref{lem1} and Condition \ref{lassp} to get $\sum_{j\in \mG}\Pr(t\leq{W}_{j}\leq t+l_n,\mC^c_t)=o(1)$.
Further note that under the event, $\{t\leq W_{j}\leq t+l_n, \mC_t\}$, we have
$$|T_{2j}|\leq \frac{t+l_n}{|T_{1j}|}\leq \frac{C(t+l_n)\sqrt{\log \bar q_n}}{t}
=C\sqrt{\log \bar q_n}+\frac{l_n\sqrt{\log \bar q_n}}{M}\leq C\sqrt{\log \bar q_n}=b_n,$$
under condition that $l_n\to 0$. Let $T_{2j}^*=\sqrt{n_2}(\widehat\mu_{2j}-\mu_j)/\sigma_j$ and $U_{j}=\sqrt{n_2}\mu_j/\sigma_j$.
Thus from Lemma \ref{klem1}, we conclude that
\begin{align*}
&\sum_{j\in \mG}\Pr(t-T_{1j}U_{j}\leq T_{1j}Z\leq t+l_n-T_{1j}U_{j}\mid \mC_t)\\
&=\sum_{j\in \mG}\E\left\{{\Phi}((t+l_n)/|T_{1j}|-U_{j})-{\Phi}(t/|T_{1j}|-U_{j}) \mid\mC_t \right\}\\
&\leq \sum_{j\in \mG}l_n\E\left\{|T_{1j}|^{-1}\phi(t/|T_{1j}|-U_{j})\mid \mC_t\right\}\\
&\leq  l_n\sum_{j\in \mG}\E\left\{\left[(t/T_{1j}^2-U_{j}/|T_{1j}|)+\frac{1}{t-U_{j}|T_{1j}|}\right]\widetilde{\Phi}(t/|T_{1j}|-U_j)\mid \mC_t\right\}\\
&\lesssim l_nM^{-1}\log \bar q_n\sum_{j\in \mG}\E\left\{\widetilde{\Phi}(t/|T_{1j}|-U_{j})\mid \mC_t\right\},
\end{align*}
where $\widetilde{\Phi}(x)=1-\Phi(x)$. The second to last inequality is due to
\[
\frac{x}{x^2+1}\phi(x)<\widetilde{\Phi}(x), \ \ \mbox{for all} \  x>0.
\]

On the other hand,
\begin{align*}
\sum_{j\in \mG}\Pr({W}_{j}>t)&=\sum_{j\in \mG}\E\left\{\widetilde{\Phi}(t/|T_{1j}|-U_j)\mid\mC_t\right\}\{1+o(1)\}.
\end{align*}
Therefore, by Markov inequality and similar arguments in the proof of Lemma \ref{klem2}, the assertion holds if $c_{np}a_{np}s_n\bar{q}_n\sqrt{n\log p}\log\bar q_nh_n\to 0$. Note that $h_n$ can be made arbitrarily small as long as $h_n\to\infty$ as $n\to\infty$, from which we completes the proof. \hfill$\Box$

In the next lemma, we obtain the approximation result for those $j$ with relatively large $\mu_j$.

\begin{lemma}\label{klemsig}
Suppose Conditions \ref{ssp}, \ref{lassp}, \ref{moment}, \ref{dm} and \ref{acc} hold and \\$c_{np}a_{np}s_n\bar{q}_n\sqrt{n\log p}(\log\bar q_n)^{1+\gamma}\to 0$. Then, for any $M>0$,
\begin{align*}
\sup_{M\leq t\leq G^{-1}(\alpha \eta_n/q_{0n})}\left|\frac{\sum_{j\in \mG^{c}}\bI(\widetilde W_{j}\geq t)}{\sum_{j\in \mG^{c}}\bI({W}_{j}\geq t)}-1\right|&=o_p(1),\\
\sup_{M\leq t\leq G_{-}^{-1}(\alpha \eta_n/q_{0n})}\left|\frac{\sum_{j\in \mG^{c}}\bI(\widetilde W_{j}\leq -t)}{\sum_{j\in \mG^{c}}\bI({W}_{j}\leq -t)}-1\right|&=o_p(1).
\end{align*}
\end{lemma}

\proof Under the designed conditions, we have $W_j=\widetilde W_j\{1+o_p(1)\}$ for any $j\in\mG^{c}$ uniformly. Then the results follow. \hfill$\Box$


\section{Proofs of Additional Theoretical Results}\label{Appendix-B}

\subsection{Proof of Lemma \ref{Delta:lem} (the coin-flip property under dependence)}

Observe that $W_j=\sqrt{n_1n_2}\widehat\mu_{1j}\widehat\mu_{2j}/\sigma^2_j=:c_j\times\widehat\mu_{2j}$. Conditional on $\D_1$, we have
$W_j\mid {\bf W}_{-j}\sim \mathcal{N}(\mu_{j|-j}, \sigma^2_{j|-j})$ with
$$\mu_{j|-j}=\cov(W_j, {\bf W}_{-j})\var({\bf W}_{-j})^{-1}({\bf W}_{-j}-\E{\bf W}_{-j}) \mbox{ and } $$
\[
\sigma^2_{j|-j}=\var(W_j)-\cov(W_j, {\bf W}_{-j})\{\var({\bf W}_{-j})\}^{-1}\cov(W_j, {\bf W}_{-j})^\top.
\]

For any $k,l\in\mS$, we have $\cov(W_k, W_l)=c_kc_lQ_{kl}$. Let ${\bf C}=\diag\{c_1,\ldots,c_{q_n}\}$ and ${\bf D}={\bf CQ C}$. Then $\cov(W_j, {\bf W}_{-j})={\bf D}_{j,-j}$ and $\var({\bf W}_{-j})={\bf D}_{-j,-j}$. So, we obtain that
\begin{align*}
&\Pr(W_j>0\mid|W_j|, {\bf W}_{-j},\D_1)\\
=&\frac{\phi\left(\frac{|W_j|-\mu_{j|-j}}{\sigma_{j|-j}}\right)}{\phi\left(\frac{|W_j|-\mu_{j|-j}}{\sigma_{j|-j}}\right)
+\phi\left(\frac{|W_j|+\mu_{j|-j}}{\sigma_{j|-j}}\right)}\\
=&\frac{\phi\left(\frac{|W_j|-{\bf D}_{j,-j}{\bf D}_{-j,-j}^{-1}({\bf W}_{-j}-\E{\bf W}_{-j})}{\sqrt{D_{jj}-{\bf D}_{j,-j}{\bf D}_{-j,-j}^{-1}{\bf D}_{-j,j}}}\right)}{\phi\left(\frac{|W_j|-{\bf D}_{j,-j}{\bf D}_{-j,-j}^{-1}({\bf W}_{-j}-\E{\bf W}_{-j})}{\sqrt{D_{jj}-{\bf D}_{j,-j}{\bf D}_{-j,-j}^{-1}{\bf D}_{-j,j}}}\right)+\phi\left(\frac{|W_j|+{\bf D}_{j,-j}{\bf D}_{-j,-j}^{-1}({\bf W}_{-j}-\E{\bf W}_{-j})}{\sqrt{D_{jj}-{\bf D}_{j,-j}{\bf D}_{-j,-j}^{-1}{\bf D}_{-j,j}}}\right)}.
\\:= & \Delta_j(|W_j|, {\bf W}_{-j},\D_1)
\end{align*}

Denote ${\bf Q}_{-j,j}=0$ the $j$th column of $\bf Q$ excluding $Q_{jj}$. Finally we have
\begin{align*}
\Pr(W_j>0\mid |W_j|, {\bf W}_{-j})&=\E\left\{\Pr(W_j>0\mid |W_j|, {\bf W}_{-j},\D_1)\mid |W_j|, {\bf W}_{-j}\right\}\\
&=\E\left\{\Delta_j(|W_j|, {\bf W}_{-j},\D_1)\mid |W_j|, {\bf W}_{-j}\right\}-1/2.
\end{align*}
It can be easily verified that if ${\bf Q}_{-j,j}=0$, $\Delta_j(|W_j|, {\bf W}_{-j},\D_1)=1/2$ and consequently $\Delta_j=0$.

\subsection{Asymptotic results for R-SDA and two--sample SDA}

The next result is a direct corollary of Theorem \ref{thm1} which establishes the FDR control of the multi-splitting procedure R-SDA.

\begin{coro}\label{coro1}
Suppose Conditions \ref{ssp}-\ref{depen} hold. For any $\alpha\in(0,1)$ and a given $B$, the FDR of the R-SDA method satisfies
$\mathop{\lim\sup}_{(n,p)\to\infty}{\rm FDR}\leq \alpha$.
\end{coro}
As in $(\ref{fa})$, the FDP is controlled for each replication so is the FDP of R-SDA, resulting in the FDR control.

To establish the FDR control result of SDA procedure for the two-sample problem, we introduce a new sequence of independent random variables $\{\bm\xi_i\}$ defined as follows:
\begin{align*}
\bm\xi_i-\bm\omega&=\left\{
         \begin{array}{c}
          n_2/n_2^{(1)}(\bm\xi^{(1)}_{2i}-\bm\mu^{(1)});\,\,\,1\leq i\leq n_2^{(1)}; \\
          -n_2/n_2^{(2)}(\bm\xi^{(2)}_{2i-n_2^{(1)}}-\bm\mu^{(2)});\,\,\,n_2^{(1)}+1\leq i\leq n_2.  \\
         \end{array}
       \right.
\end{align*}

Note that
\begin{align*}
{\bar{\bm\xi}}^{(1)}_2-{\bar{\bm\xi}}^{(2)}_2-\bm\omega
=\frac{1}{n_2^{(1)}}\sum_{i=1}^{n_2^{(1)}}(\bm\xi^{(1)}_{2i}-\bm\mu^{(1)})-\frac{1}{n_2^{(2)}}\sum_{i=1}^{n_2^{(2)}}(\bm\xi^{(2)}_{2i}-\bm\mu^{(2)})
=\frac{1}{n_2}\sum_{i=1}^{n_2}(\bm\xi_i-\bm\omega).
\end{align*}

By the proofs for Theorem \ref{thm1}, if we replace $\bm\mu$ as $\bm\omega$ and set $\bm\Omega^{-1}=\bm\Sigma^{(1)}/\varrho+\bm\Sigma^{(2)}/(1-\varrho)$ with $\varrho=\lim n^{(1)}_l/n_l$, Theorem \ref{thm1} holds also for the two-sample problem.

\begin{coro}\label{coro2} Suppose Conditions \ref{ssp}-\ref{depen} hold. For any $\alpha\in(0,1)$ and $0<\varrho<1$, the FDR of the SDA for the two-sample problem satisfies
$\mathop{\lim\sup}_{(n,p)\to\infty}{\rm FDR}\leq \alpha$.
\end{coro}

We want to emphasize that as long as Condition \ref{lassp} is satisfied, the above results hold for other choices of $T_{1j}$ as discussed in Appendix~\ref{other-t1.sec}. For example, consider a hard-thresholding estimator $\widehat{\mu}_{1j}=\bar{\xi}_{1j}\bI(|\bar{\xi}_{1j}|>c\sqrt{\log p/n})$ for some $c>0$. We know that $c_{np}=\sqrt{\log p/n}$ if  $\xi_{ij}$'s have uniformly bounded fourth moments.

\subsection{Additional lemmas}

The first one is the standard Bernstein's inequality.
\begin{lemma}[Bernstein's inequality]\label{lemma:bernstein}
Let $X_1,\ldots,X_n$ be independent centered random variables a.s. bounded by $A<\infty$ in absolute value. Let $\sigma^2=n^{-1}\sum_{i=1}^n\E(X_i^2)$. Then for all $x>0$,
\begin{eqnarray*}
  \Pr\Big(\sum_{i=1}^n X_i\geq x\Big)\leq\exp\Big(-\frac{x^2}{2n\sigma^2+2Ax/3}\Big).
\end{eqnarray*}
\end{lemma}

The second one is a moderate deviation result for the mean; See \cite{petrov2002probabilities}.
\begin{lemma}[Moderate deviation for the independent sum]\label{mdm}
Suppose that $X_1,\ldots,X_n$ are independent random variables with mean zero, satisfying $\E(|X_j|^{2+\delta})<\infty$ ($j=1,2,\ldots$). Let $B_n=\sum_{i=1}^n\E(X_i^2)$. Then,
\[
\frac{\Pr(\sum_{i=1}^nX_i>x\sqrt{B_n})}{1-\Phi(x)}\to 1,
\]
as $n\to\infty$ uniformly in $x$ in the domain $0\leq x\leq C\{2\log(1/L_n)\}^{1/2}$, where $L_n=B_n^{-1-\delta/2}\sum_{i=1}^n\E|X_i|^{2+\delta}$ and $C$ is a positive
constant satisfying the condition $C<1$.
\end{lemma}

The next lemma establishes uniform bounds for $\widehat{\mu}_{2j}$.
\begin{lemma}\label{lem1} Suppose Conditions \ref{ssp}, \ref{moment}, and \ref{dm} hold. Then, as $n\to\infty$,
\begin{align*}
\Pr\left(\sigma_{j}^{-1}\left|\widehat{\mu}_{2j}-\mu_{j}\right|>\sigma\sqrt{C\log \bar{q}_n/n_2}\mid \D_1\right)=o(1/\bar{q}_n),
\end{align*}
holds uniformly in $\mS$, where $C>4$.
\end{lemma}

\proof
Write
\begin{align*}
&\widehat{\mu}_{2j}-\mu_{j}=n^{-1}_2\sum_{i=1}^{n_2}\e^\top_j\left(\X^\top_{2\mS}\X_{2\mS}\right)^{-1}\X_{2\mS}^\top \bm\varepsilon_i:=n^{-1}_2\sum_{i=1}^{n_2}\epsilon_{ij}.
\end{align*}
Let $m_n=(n_2\bar{q}_n)^{1/\theta+\gamma}K_{n2}$ and note that
\begin{align*}
&\epsilon_{ij}=\epsilon_{ij}\mathbb{I}(|\epsilon_{ij}|\leq m_n)-\mathbb \E\{\epsilon_{j}\mathbb{I}(|\epsilon_{j}|\leq m_n)\}
+\epsilon_{ij}\mathbb{I}(|\epsilon_{ij}|> m_n)-\mathbb \E\{\epsilon_{j}\mathbb{I}(|\epsilon_{j}|>m_n)\}\\
&=:\epsilon_{ij,1}+\epsilon_{ij,2}.
\end{align*}

Conditioned on the first split $\D_1$,
\begin{align}\label{Punif}
&\Pr\left(\left|\sqrt{n_2}\left(\widehat{\mu}_{2j}-\mu_j\right)\right|>\sigma_jx\ \ \mbox{for some}\ j\mid \D_1 \right)\nonumber\\
&=\Pr\left(\left|\sum\limits_{i=1}^{n_2}{\epsilon}_{ij,1}+\sum\limits_{i=1}^{n_2}{\epsilon}_{ij,2}\right|>\sqrt{n_2}\sigma_{j}x\ \ \mbox{for some}\ j\mid \D_1 \right)\nonumber\\
&\leq \Pr\left(\left|\sum\limits_{i=1}^{n_2}{\epsilon}_{ij,1}\right|+\left|\sum\limits_{i=1}^{n_2}{\epsilon}_{ij,2}\right|>\sqrt{n_2}\sigma_{j}x \ \ \mbox{for some}\ j\mid \D_1 \right)\nonumber\\
&\leq \Pr\left(\left|\sum\limits_{i=1}^{n_2}{\epsilon}_{ij,1}\right|>\sqrt{n_2}\sigma_{j}x(1-a) \ \ \mbox{for some}\ j\mid \D_1 \right)\nonumber\\
&+\Pr\left(\left|\sum\limits_{i=1}^{n_2}{\epsilon}_{ij,2}\right|>\sqrt{n_2}\sigma_{j}xa \ \ \mbox{for some}\ j\mid \D_1 \right)=:P_1+P_2.
\end{align}
Here $a$ is a small positive value.


Firstly consider the term $P_1$. Note that $\epsilon_{1j,1},\ldots,\epsilon_{n_2j,1}$ are independent centered random variables a.s. bounded by $2m_n$ in absolute value. Then the Bernstein inequality in Lemma \ref{lemma:bernstein} yields that
\begin{align*}
P_1&\leq 2q_n\max_j\exp\left\{-\frac{n_2\sigma_{j}^2x^2(1-a)^2}{2n_2\mathbb E(\epsilon_{j,1}^2)+2\cdot 2m_n\cdot\sqrt{n_2}\sigma_{j}x(1-a)/3}\right\}.
\end{align*}
Recall that $\epsilon_{ij,1}=\epsilon_{ij}\mathbb{I}(|\epsilon_{ij}|\leq m_n)-\mathbb E[\epsilon_{j}\mathbb{I}(|\epsilon_{j}|\leq m_n)]$.
Thus
\begin{align*}
\mathbb \E(\epsilon_{j,1}^2)=\mbox{Var}\{\epsilon_{j}\mathbb{I}(|\epsilon_{j}|\leq m_n)\}\leq \mathbb \E\{\epsilon_{j}^2\mathbb{I}(|\epsilon_{j}|\leq m_n)\}
\leq  \mathbb \E(\epsilon_{j}^2)=Q_{jj}.
\end{align*}
We then have:
\begin{align}\label{P1}
P_1&\leq 2q_n\max_j\exp\left\{-\frac{n_2\sigma_{j}^2x^2(1-a)^2}{2n_2Q_{jj}+2\cdot 2m_n\cdot\sqrt{n_2}\sigma_{j}x(1-a)/3}\right\}\nonumber\\
&\leq 2\bar{q}_n\max_j\exp\left\{-\frac{x^2(1-a)^2}{2\sigma^2+4(1-a)\sigma_{j}^{-1}xm_n/(3\sqrt{n_2})}\right\}.
\end{align}

Next we turn to consider $P_2$. First note that
\begin{align*}
P_2&\leq \Pr\left(\sum\limits_{i=1}^{n_2}\max_j|\epsilon_{ij}|\mathbb{I}(|\epsilon_{ij}|>m_n)+\max_j n_2\mathbb E\{|\epsilon_{j}|\mathbb{I}(|\epsilon_{j}|>m_n)\}>\sqrt{n_2}\sigma_{j}xa\mid \D_1 \right)
\end{align*}

Further note that
\begin{align*}
\mathbb E^2\{|\epsilon_{j}|\mathbb{I}(|\epsilon_{j}|>m_n)\}\leq \mathbb E(\epsilon_{j}^2)\Pr(|\epsilon_{j}|>m_n)\leq
\mathbb E(\epsilon_{j}^2)\frac{\mathbb E(|\epsilon_{j}|^{\theta})}{m_n^{\theta}}.
\end{align*}
We then conclude that
\begin{align*}
\max_jn_2\mathbb E\{|\epsilon_{j}|\mathbb{I}(|\epsilon_{j}|>m_n)\}\leq \max_jn_2\frac{\sqrt{\mathbb E(\epsilon_{j}^2)\mathbb E(|\epsilon_{j}|^{\theta})}}{m_n^{\theta/2}}=o(\sqrt{n_2}).
\end{align*}
From this, we then have
\begin{align}\label{P2}
P_2&\leq \Pr\left(\sum\limits_{i=1}^{n_2}\max_j|\epsilon_{ij}|\mathbb{I}(|\epsilon_{ij}|>m_n)>\sqrt{n_2}\sigma_{j}xa/2\mid \D_1 \right)\nonumber\\
&\leq \Pr\left(\max_j|\epsilon_{ij}|> m_n\ \ \mbox{for some}\ i\mid \D_1 \right)\nonumber\\
&\leq n_2\frac{\mathbb E(\|{\bf A}(\mS)\bm\varepsilon_{i}\|_{\infty}^{\theta})}{m_n^{\theta}}=o(\bar{q}_n^{-1}).
\end{align}

Let $x=\sigma\sqrt{C\log \bar{q}_n}$. From the inequalities (\ref{Punif}), (\ref{P1}), and (\ref{P2}), we conclude that
\begin{align*}
&\Pr\left(\left|\sqrt{n_2}\left(\widehat{\mu}_{2j}-\mu_j\right)\right|>\sigma_jx\ \ \mbox{for some}\ j\mid \D_1 \right)\nonumber\\
&\leq 2\bar{q}_n\max_j\exp\left\{-\frac{x^2(1-a)^2}{2\sigma^2+4(1-a)\sigma_{j}^{-1}xm_n/(3\sqrt{n_2})}\right\}+o(\bar{q}_n^{-1})
=o(\bar{q}_n^{-1}).
\end{align*}
holds uniformly in $\mS$, where we use the condition $m_n/\sqrt{n/\log \bar{q}_n}=o(1)$ which is implied by Condition \ref{moment}.  \hfill$\Box$

\section{Additional Numerical Results}\label{Appendix-C}

\subsection{Estimated covariance structures} \label{est-cov:sec}

This section compares the methods mentioned in Section~\ref{Sec-4} for the unknown covariance case. 
In practice, one should adopt the most appropriate estimator tailored to specific correlation structures. Specifically, we have used the method based on Cholesky decomposition in \cite{bickel2008regularized}, the POET method proposed by \cite{fan2013large}, and the graphical lasso \citep{friedman2008sparse} to estimate the unknown Structures (I)--(III), respectively.

\begin{figure}
  \centering
  \includegraphics[width=0.9\textwidth]{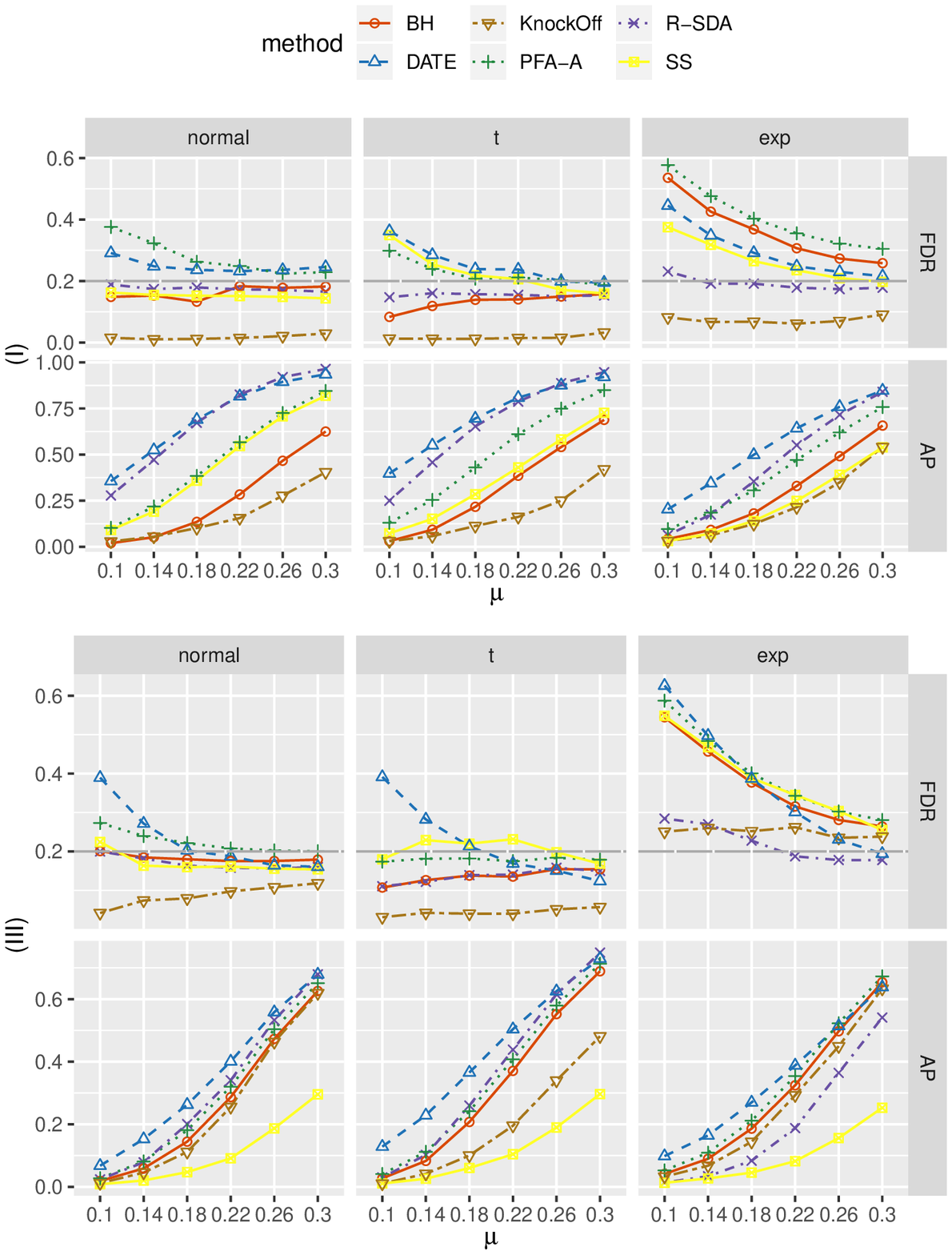}
\vspace{-0.25in}
  \caption{\small\it FDR and AP comparison for varying $\mu$ in Settings (I) and (III) with estimated covariance matrix.}
  \label{Fig-mu-est}
\end{figure}

\begin{figure}[ht]
  \centering
  \includegraphics[scale=0.8]{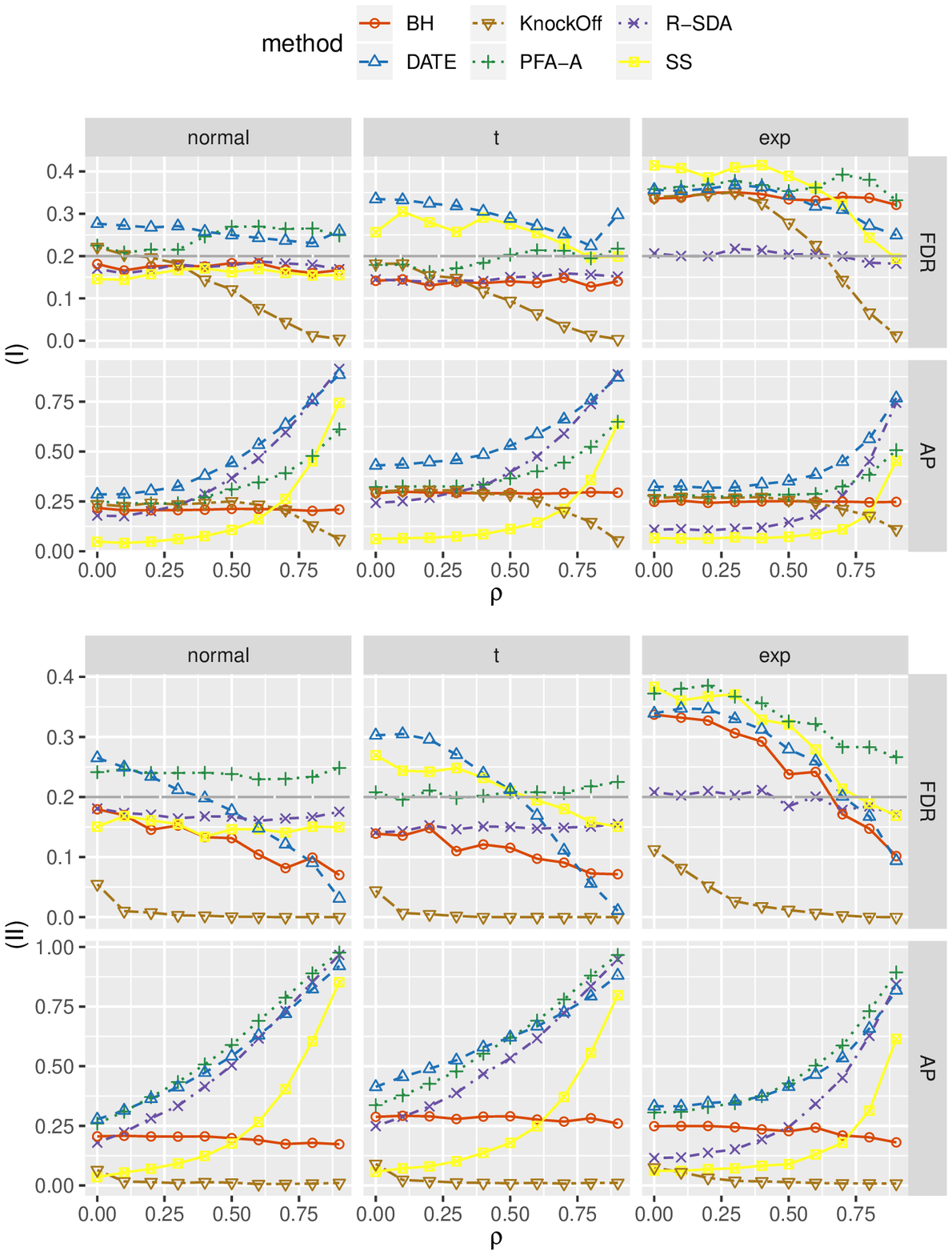}
\vspace{-0.25in}
  \caption{\small\it FDR and AP comparison for varying $\rho$ in Settings (I)--(II) with estimated covariance matrix.}
\label{Fig-rho-est}
\end{figure}

Figure~\ref{Fig-mu-est} follows the settings in Figure~\ref{Fig-mu} (except that the covariance matrix or its inverse is estimated).
Figures~\ref{Fig-rho-est} uses the same settings as those in Figures~\ref{Fig-rho} with estimated covariance matrix.
We omit a detailed discussion as the observed patterns seem to be very similar to those in the known {covariance} case (except that the FDR control sometimes becomes less accurate due to the additional estimation errors). Our conclusions based on Figure~\ref{Fig-mu-est} and Figures~\ref{Fig-rho-est} remain essentially the same as before. Knockoff and R-SDA seem to be the only methods that can control the FDR reasonably well in all scenarios, with the R-SDA method having much higher power in most scenarios.


\subsection{Additional comparisons}
\begin{figure}[ht]
  \centering
  \includegraphics[scale=0.8]{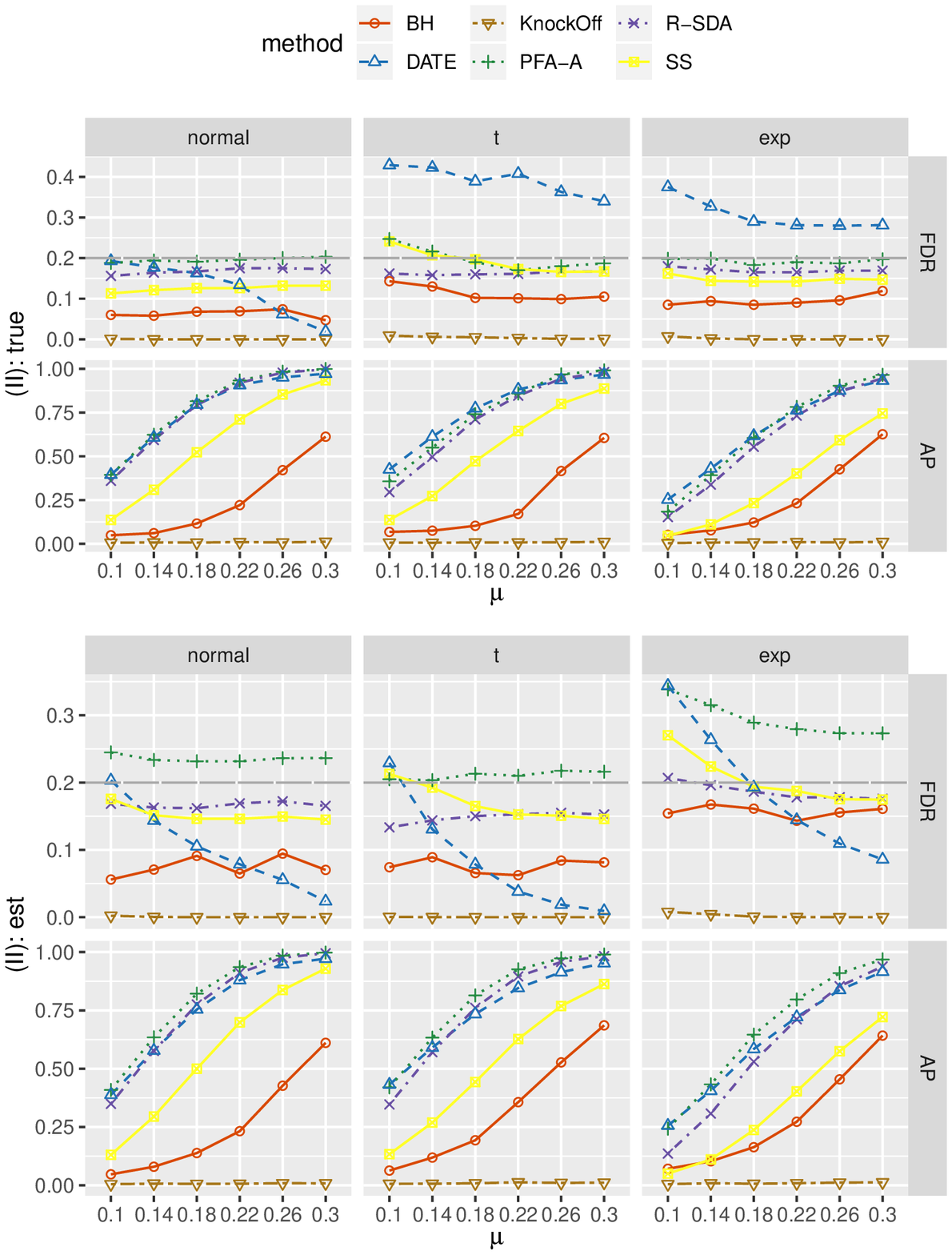}
\vspace{-0.25in}
  \caption{\small\it FDR and AP comparison for varying $\mu$ in Setting (II) with known (top half) and unknown variances (bottom half).}\label{Fig-II}
\end{figure}

Figure~\ref{Fig-II} demonstrates the FDR and AP for various signal magnitude $\mu$ under the compound symmetry error structure (II) and three error distributions, for known and unknown covariance structures, respectively.

\subsection{Boxplots of FDPs}
\begin{figure}[ht]
  \centering
  \includegraphics[scale=0.55]{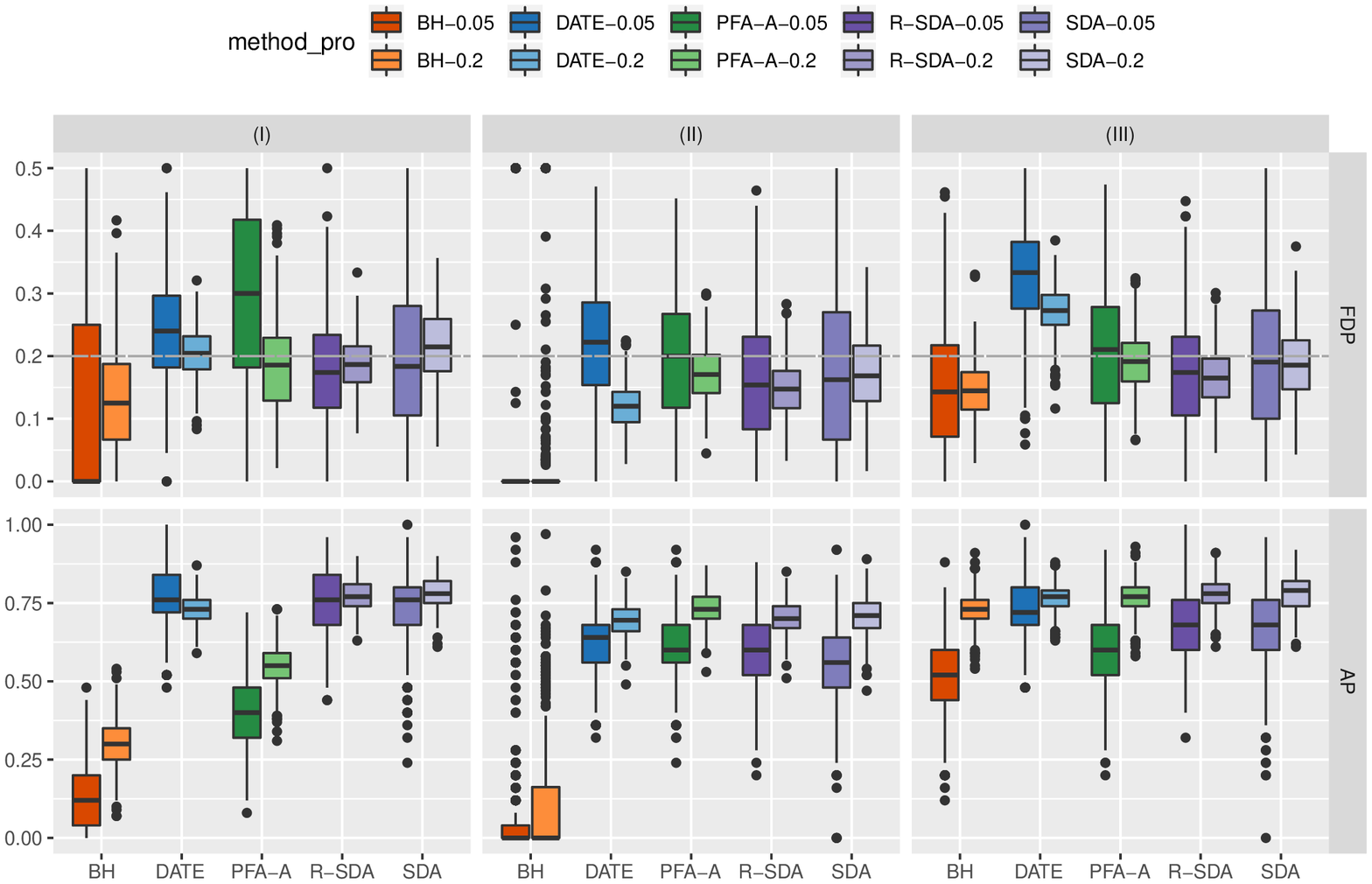}
  \caption{\it\small The boxplot of FDP and AP when the proportions of alternative are 0.05 and 0.2. The normal error is considered and $(n, p, \alpha)=(90, 500, 0.2)$. The signal strength $\mu$ is set as 0.2, 0.15, 0.3 for the covariance structures (I)-(III), respectively.}\label{Fig-1}
\end{figure}

When the noises are sampled from the multivariate normal distribution, Figure~\ref{Fig-1} shows the boxplot of the FDP and AP of the testing procedures for $\pi_1=0.05$ and 0.2, while fixing $(n, p, \alpha)=(90, 500, 0.2)$. The signal magnitude $\mu$ is adjusted according to the covariance structures so that the APs are in a similar range. While the BH is conservative with little power, the R-SDA outperforms the DATE and PFA in the sense that it provides more accurate estimate of FDP and generally higher power. The conclusions are consistent for different choices of $\pi_1$, with narrower interquartile range of FDP and AP for larger $\pi_1$. As we can expect, the R-SDA has smaller variation than the single-splitting SDA.

\subsection{The impact of the number of tests and sample sizes}
\begin{figure}[ht]
  \centering
  \includegraphics[width=5.1in]{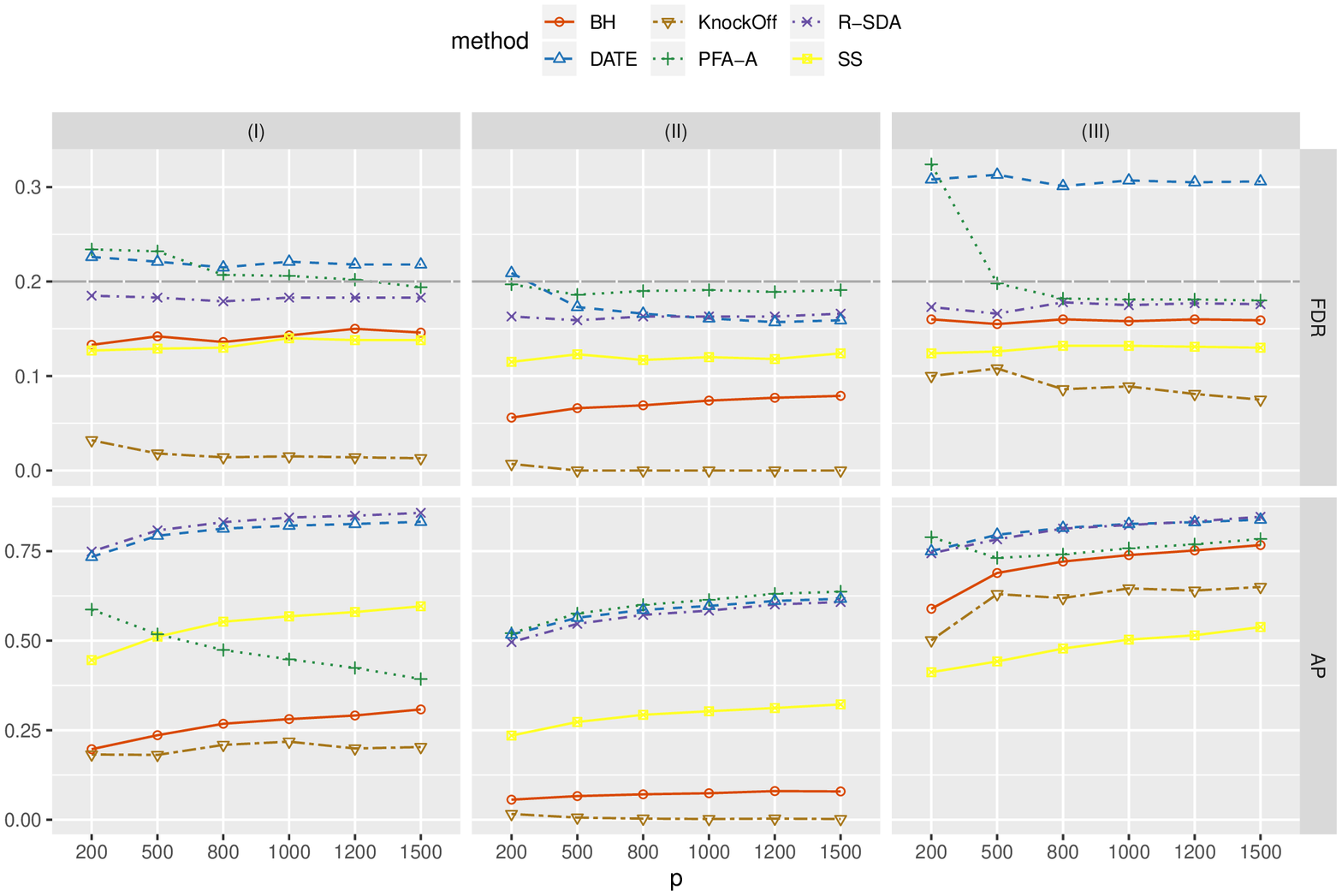}
    \includegraphics[width=5.1in]{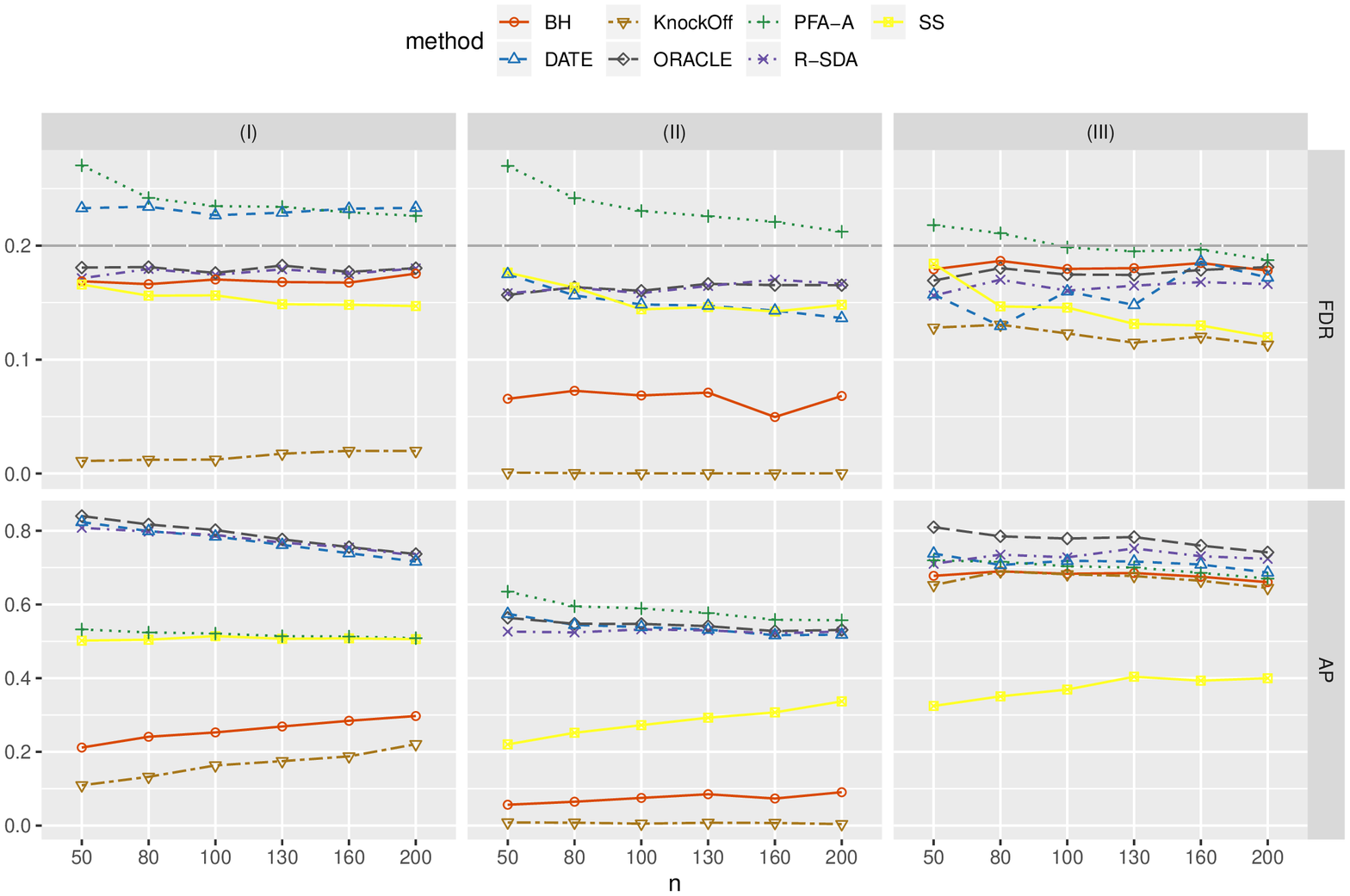}
  \caption{\small\it Top half: The empirical FDR and AP for varying $p$. $(n, \pi_1, \alpha)=(90, 0.1, 0.2)$ and $\mu_n=C\sqrt{\log(p)/n}$ with $C=0.8, 0.5, 1.2$. Bottom half: The FDR and AP for varying $n$ when the covariances are estimated. $(p, \pi_1, \alpha)=(500, 0.1, 0.2)$ and $\mu_n=C\sqrt{\log p/n}$ with $C=0.8, 0.5, 1.2$.}\label{Fig-3}
\end{figure}

We also conduct experiments by altering the number of tests $p$, while keeping $(n, \pi_1, \alpha)=(90, 0.1, 0.2)$. To make the AP comparable across $p$,  the signal $\mu$ is adjusted via $\mu = C\sqrt{\log p/n}$ with $C$ depending on the covariance structures.  The results are summarized in the top half of Figure \ref{Fig-3}. We can see that all methods have more accurate control of FDR as $p$ increases, but the $\mathrm{PFA}_{\mathrm{A}}$ and DATE fail to control the FDR when $p$ is small. To investigate the impact on sample sizes with unknown covariance, we set $\mu=C\sqrt{\log(p)/n}$, fix $(p, \pi_1, \alpha)=(500, 0.1, 0.2)$, and consider the normal error. The results are summarized in the bottom half of Figure~\ref{Fig-3}. We can see that that our R-SDA method is able to control the FDR and close to  the nominal level regardless of the choice of $n$. Its superior performance relative to the other three methods is significant in some cases. Though all the methods exhibit steady AP pattern, the BH, PFA and DATE appear to need larger sample to achieve satisfactory FDR control than the R-SDA does. This again concurs with our theoretical result in Theorem \ref{thm1} and demonstrates the advantage of using the nonparametric estimation of FDP in the SDA procedure.

\subsection{List of selected genes by different methods}

Table \ref{Table-genes} reports the list of 19 most differentially expressed
probe sets obtained by the methods R-SDA, BH, SS, PFA-A and DATE in the real-data example.
{\small
\begin{table}[ht] 
\begin{scriptsize}
\caption{\textsl{\small Differentially expressed probe sets in the B lineage \textsl{ALL} with BCR/ABL versus NEG molecular rearrangement, for five different multiple testing adjustment methods}}  \label{Table-genes}
\begin{center}  \tabcolsep 18.5pt
\begin{tabular}{lllll}  \hline
R-SDA	&	BH	&	SS	&	PFA	&	DATE	\\ \hline
1635$\_$at	&	1636$\_$g$\_$at	&	39730$\_$at	&	1636$\_$g$\_$at	&	36502$\_$at	\\
39730$\_$at	&	39730$\_$at	&	39317$\_$at	&	39730$\_$at	&	38385$\_$at	\\
1636$\_$g$\_$at	&	1635$\_$at	&	37027$\_$at	&	1635$\_$at	&	40202$\_$at	\\
36502$\_$at	&	1674$\_$at	&	38052$\_$at	&	1674$\_$at	&	37403$\_$at	\\
37403$\_$at	&	40504$\_$at	&	1635$\_$at	&	40202$\_$at	&	38052$\_$at	\\
32134$\_$at	&	40202$\_$at	&	1636$\_$g$\_$at	&	37403$\_$at	&	33690$\_$at	\\
38052$\_$at	&	37015$\_$at	&	40202$\_$at	&	32434$\_$at	&	39317$\_$at	\\
36821$\_$at	&	37027$\_$at	&	34850$\_$at	&	37014$\_$at	&	40876$\_$at	\\
38385$\_$at	&	32434$\_$at	&	37403$\_$at	&	32979$\_$at	&	33440$\_$at	\\
37027$\_$at	&	40167$\_$s$\_$at	&	37024$\_$at	&	1249$\_$at	&	1674$\_$at	\\
1674$\_$at	&	40480$\_$s$\_$at	&	1249$\_$at	&	38111$\_$at	&	36908$\_$at	\\
41872$\_$at	&	36591$\_$at	&	36802$\_$at	&	37015$\_$at	&	33774$\_$at	\\
33440$\_$at	&	33774$\_$at	&	37025$\_$at	&	37147$\_$at	&	39730$\_$at	\\
32434$\_$at	&	37403$\_$at	&	32979$\_$at	&	40504$\_$at	&	41592$\_$at	\\
40876$\_$at	&	37014$\_$at	&	34870$\_$at	&	33440$\_$at	&	32134$\_$at	\\
40202$\_$at	&	37363$\_$at	&	36502$\_$at	&	38112$\_$g$\_$at	&	39070$\_$at	\\
39317$\_$at	&	34472$\_$at	&	33891$\_$at	&	36502$\_$at	&	37558$\_$at	\\
32562$\_$at	&	32542$\_$at	&	34800$\_$at	&	31786$\_$at	&	33304$\_$at	\\
34990$\_$at	&	39329$\_$at	&	36543$\_$at	&	34850$\_$at	&	34180$\_$at	\\ \hline
\end{tabular}
\end{center}
\end{scriptsize}
\end{table}}

\end{document}